\documentclass[aip,prf,reprint,groupedaddress,onecolumn]{revtex4-2}%
\usepackage[usenames,dvipsnames]{color}
\usepackage{subeqnarray}
\usepackage{amsmath}
\usepackage[disable]{todonotes}
\usepackage{graphicx}
\usepackage{natbib}
\usepackage{wasysym}
\usepackage{amssymb}
\usepackage{xcolor}
\usepackage{tikz} 
\usetikzlibrary{positioning}
\usepackage{amsthm}
\usepackage{moreverb} 
\usepackage{textcomp} 
\usepackage{epsfig}
\usepackage{color} 
\usepackage{upgreek}
\usepackage{bbm}
\usepackage{multirow}
\usepackage{graphics}
\usepackage{dcolumn}
\usepackage{bm}
\usepackage{amsfonts}
\usepackage{textcomp} 
\usepackage{caption}
\usepackage{subcaption}

\begin{document}

\preprint{APS/123-QED}

\title{
Prediction of resistance induced by surface complexity in lubricating layers:  Application to super-hydrophobic surfaces}

\author{Noura Bettaieb}
\author{Marco Castagna}
\author{Pierre-Yves Passaggia}
\author{Azeddine Kourta}
\author{Nicolas Mazellier}%
\email{nicolas.mazellier@univ-orleans.fr}
\affiliation{%
University of Orl\'eans, INSA-CVL, PRISME, EA 4229, 45072, Orl\'eans, France 
}%




\date{\today}

\begin{abstract}
Super Hydrophobic (SH) coatings are widely used to mitigate drag in various applications. Numerous studies have demonstrated that the beneficial wall-slip effect produced by these materials disappears in laminar flow regimes. The main mechanisms considered to be behind the decrease in performance are Marangoni-induced stresses and air/liquid interface deformation. In the present study, a new mechanism is proposed to explain the loss of performances of SH-surfaces in laminar flow regimes. Here we consider the flow of air inside the plastron and the associated momentum loses induced by roughness elements with different geometric characteristics. The effects of air motion within the plastron is coupled to the outer fluid with a homogenised boundary condition approach. To this end, numerical simulations at the scale of the roughness element were conducted as a function of the porosity and the tortuosity of the domain to determine the slip velocity at the air-liquid interface.
The homogenised boundary condition is then implemented in a theoretical model for the outer flow to compute drag on SH-spheres at low $Re$ numbers. Experiments of laminar SH falling spheres indicate that high values of the tortuosity and low values of porosity lead to a loss of performances when considering drag reduction. As anticipated, a 3D printed sphere with low tortuosity and similar porosity demonstrated near-optimal drag reductions. A comparative study between the predicted values and experiments shows that the homogenised model is able to accurately predict the drag on SH surfaces for values of the porosity and tortuosity estimated from microscopy images of the SH textured surface.

\end{abstract}

\maketitle

\section{Introduction}

More than a century after the seminal work of Stokes \cite{Stokes1851}, bluff-body drag and in particular its control remains an open question. Combined efforts from both theoretical fluid dynamics \cite{Albano1975} and material sciences \cite{Quere2008} demonstrated the potential of wall-slip-type surfaces towards drag reduction. Ideally, gas entrapment between the liquid and the solid induces a local slip condition which may favourably decrease drag via momentum transfer at the gas/liquid interface \cite{McHale2011}. Gas lubrication can be practically achieved by using methods such as gas injection \cite{ceccio2010friction}, Leidenfrost effect \cite{vakarelski2011drag}, or combining surface texturing and chemical repellency \cite{vinogradova1999}. The latter approach leads to the so-called super-hydrophobic (SH) surfaces \cite{Rothstein2010}, which are at the core of this study.

The SH property of the surface generates a high contact angle that repels water when in contact with the surface. Moreover, the presence of roughness elements on the surface results in cavities that entrap air and prevent water from infusing. These characteristics explain the potential of SH surfaces of reducing hydrodynamic drag. Precisely, since the surface retains a plastron, also referred to as a lubricating air layer, the friction resistance is decreased. The presence of an air layer changes the boundary condition at the wall from a zero-velocity condition to an interface between water and the entrapped air. 
 
Under these circumstances, the fluid experiences a sufficient slip on the surface that may reduce drag for both low and high Reynolds number flow \cite{Gruncell2013}. Nevertheless, regarding flows over bluff-bodies, few available experimental studies \cite {Byon2010, Ahmmed2016, modak2017creeping} report mild-to-negligible drag reduction unlike predicted by theoretical and numerical works based on an ideal gas/liquid interface \cite{Gruncell2013,Peaudecerf2017}. For instance, Ahmmed et al. \cite{Ahmmed2016} showed experimentally that SH surfaces may not contribute significantly to drag reduction in laminar flows. Precisely, they found that SH spheres presented higher drag compared to non-SH spheres. They concluded that although SH surfaces may reduce the friction drag, the form drag may increase due to the texture of the water repelling surfaces. Whence, they assumed that the increase in the form drag may outweigh the resultant SH effect, hence the overall drag could increase. 
Overall, these contradictions emphasise the need for improving the physical modelling of SH surfaces. One important aspect towards a better modelling would consist in identifying the mechanism behind performance degradation of SH surfaces.

\begin{figure}[t]
\centerline{\scalebox{0.4}{\huge\input{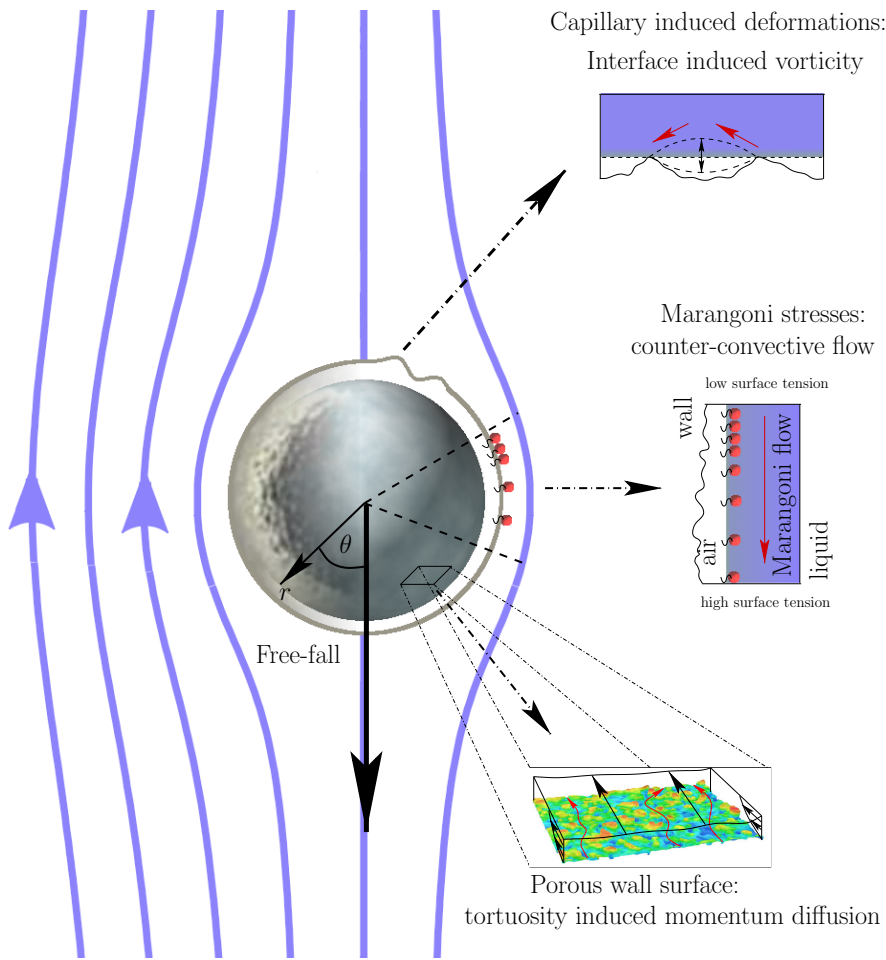}}}%
\caption{Schematics summarising the possible mechanisms influencing the drag-reduction capabilities
of SH coatings.}
\label{fig:schematic}
\end{figure}

Previous studies attempted to associate this mechanism considering various effects. Fig.\ref{fig:schematic} illustrates some of the common effects that may contribute to degrading the performance of SH-surfaces. Peaudecerf et al. \cite{Peaudecerf2017} studied the Marangoni flow generated as a result of the surface tension gradient due to the build-up of contaminants on the air/liquid interface. They showed that surfactant-induced stresses can become significant, even for very low contaminants concentrations, potentially yielding a no-slip boundary condition over the flat air/liquid interface. These results were further extended by Song et al. \cite{Song2018} who demonstrated that Marangoni effects are dependent on the roughness arrangement. Comparing closed cavities and continuous grooves, they showed that preventing surfactant to build-up in the latter case induces a negligible surface tension gradient, restricting thereby the adverse effect of the Marangoni flow.
Landel et al.\cite{LandelSurfactantModel} proposed a model to improve surfactant predictions for slip velocities. Their study was based on a model for two-dimensional, laminar, steady and pressure driven flow in a periodic SH channel. They showed how drag and slip depend on the characteristics of surfactant transport near the SH surface.

The deformation of the air-liquid interface may have an impact on the hydrodynamic performance of SH-surfaces. Castagna et al.\cite{Castagna2021,Castagna2018} investigated  experimentally the role of air plastron deformation on the performance of falling SH-spheres for different $Re$ flow regimes. They demonstrated that the local distortion of the air pocket can trigger an early transition in the wake. More accurately, their results confirm that a slip Boundary Condition (BC) on the wall of the sphere may cause wake transition earlier than a no-slip BC causing the loss of the slippery effect.
Additionally, they found that surface texture has an indispensable role on the value of the critical $Re$ that triggers wake instabilities. Kim and Rothstein\cite{Kim-Rothstein2017} conducted a set of experiments to study the effect of composite interface shape on laminar-drag reduction. Velocity and pressure drop measurements were conducted across a regularly distributed array of SH-pillars. The shape of the contact surface between the outer liquid and air enclosed inside the SH-apple-core-shaped pillars was systematically altered from concave to convex. Results indicated that the intensity of the slip velocity strongly depends on the shape of the air-water interface. Further, the slip length was found to decrease as the interface grew from convex to flat to concave. Moreover, they highlighted that the maximum tangential velocity was found to be $45\%$ of the average streamwise velocity. These conclusions were also presented in the work of Song et al. \cite{Song2018}. In addition to their investigation of the Marangoni effects, they observed that the flow over closed cavities is sensitive to the shape of the air/liquid interface. Convex air/liquid interfaces were found to maximise the slip velocity with respect to the concave counterpart.

Another mechanism (not portrayed in Fig.\ref{fig:schematic}) may also be at stake for explaining the decrease in SH-surface efficiency. The fragmentation of the pocket air into the surrounding liquid can result in a decrease of the effective slip length. Govardhan et al.\cite{Govardhan2009} studied experimentally the flow across randomly textured hydrophobic surfaces with water as the working fluid. Direct visualisations of enclosed air pockets were performed for the purpose of understanding the time dependent slip length in stationary flows. The visualisation of the bright spots at the composite interface, which were based on the principle of total internal reflection of light, confirmed the decrease of the number of air pockets with time. Henceforth, the time scale for the decrease in the slip length would eventually control the time for which the SH-surface remains efficient. Another study by Ling et al.\cite{Ling2017} studied the impact of gas diffusion on the volume of the plastron. Factors such as the flow regime and the ambient pressure were considered. Experiments demonstrated that the pressure increase resulted in a migration of the interface into the groove, together with a high advancing contact angle. Whereas, a pressure decrease caused an upward migration of the contact surface resulting in a convex interface. It was also found that the diffusion rate increases with increasing $Re$. 

Previous works confirm that the features of the plastron has an impact on the performance of SH-surfaces. In other words, the air flow inside the plastron may have a key role in determining the efficiency of the water repellent surface. This effect has been studied numerically by  Gruncell et al. \cite{Gruncell2013} where baffles, used to model roughness elements, resulted in a flow with a recirculation region  unlike the case of perfect plastron. 
Consequently, the drag reduction effect decreased since the slip velocity around the edges of the baffles (same behaviour for real roughness elements) also decreased. Furthermore, an increase in the Solid Fraction (SF) of the textured surface reduces the performance of SH-surfaces. To be exact, the decrease in the porosity of the domain, caused by the increase of SF, may generate an increase in the relative blockage, hence, a drag increase. Indeed, the porosity of the plastron has a great impact on the performance of SH surfaces and should therefore be considered when quantifying the slippage effect of SH surfaces. Regarding quantifying the super-lubricating effect, previous studies already proposed scaling laws that depends on the generic geometric characteristics of the surface. Ybert {\it et al.}\cite{ybert2007} proposed expressions that can be used to determine frictional properties of SH surfaces. The solid fraction of the interface, roughness length scale, among others where found to play a major role. 

In this work, we take another step at modelling the role of textured SH surfaces in modifying friction drag.
In particular, a novel effect responsible for reducing the efficiency of SH-surfaces is determined. Combining the air flow features inside the plastron in addition to the surface texture characteristics, we show that the increase skin-friction can be related to the movement of air between the roughness elements, inside the air layer. Specifically, we propose a simple model for the interaction of the moving air within roughness elements with complex patterns. The motion in the air layer is found to induce enough friction, capable of affecting slip at the gas-liquid interface. It is quite obvious that the porosity of the domain has a key role in determining the intensity of the drag. A larger domain porosity yields a larger interface between the liquid and the gas layer, hence a larger slip velocity. At the contrary, it is not pronounced that the path length the air is forced to follow may have an impact on the amount of liquid slippage. It is worth noting that the air flow path, also referred to as the tortuosity, have not been considered in any of the previous cited research.

Fig.~\ref{fig:schematic} illustrates the idea that air pockets are encapsulated between roughness elements, similarly to a porous medium. In that case, a porous medium approach can be used to model the motion of air through the surface texture. In other words, instead of solving the fine scale details at the interface level, one can implement a single boundary condition, capable of describing the slip velocity of the fluid considering only geometric features such as the tortuosity and the porosity. 

A simple yet effective way to ascribe a complex BC can be performed using homogenisation approaches. 
For instance, Naqvi and Bottaro\cite{Sahrish-Alessandro2021} explored the interfacial conditions between a free fluid and a porous medium based on a Beavers-Joseph-Saffman condition. The study considered an isotropic medium approach through which, model coefficient were derived for two-dimensional- and three-dimensional-type textures. The proposed condition was verified through feature resolving simulations up to higher Reynolds number compared to those valid for the classical theory. In another study, Bottaro and Naqvi\cite{Alessandro2020} proposed two different approaches to develop effective boundary conditions that model flow over walls with regularly distributed micro-structures. The method was based on harmonising the solution of the outer flow dependent on macroscopic variables to that of the inner flow that depends on large and small scale variables. The effect of surface texture is transferred to the outer flow through the inner equations at the interface level. 
The same concept is considered in this study in order to determine a homogenised BC that determines slip velocities for a SH material, in terms of the porosity and tortuosity of the medium. In this case, the effect of the roughness elements arrangement on the dynamics of the outer flow can be captured through the homogenised slip velocity expressions. A relationship between drag in creeping flow and slip velocity is thereafter needed.

The paper is organised as follows: In \S II, Stokes' flow around a sphere with an imposed slip velocity at the wall is first proposed, followed by a homogenisation of the boundary condition at the air/liquid. The BC at the composite interface is expressed in terms of the geometric characteristics of the surface (porosity and tortuosity) which are determined via numerical investigations. In \S III, the falling experiments of SH-spheres with random distributed roughness elements are introduced. The performance of spheres with different coatings is first discussed in \S IV. A comparison of the predicted drag using the proposed model and that obtained from experiments is performed. In the light of the obtained results, an additional SH-sphere with structured roughness elements is manufactured and tested both experimentally and theoretically. Conclusions are drawn in \S VI.

\section{Stokes' flow around a sphere with partial slip}

The homogenisation procedure begins with solving the laminar flow around a sphere with a slip velocity at the surface of the sphere. The value of the slip velocity are determined by the characteristics of the SH surface in \S \ref{subsec:homogenised} and drag is used as a global measure for the validation with experiments. 

\subsection{Creeping flow around a SH-sphere}

In the present model, the tangential component of the velocity is defined as a function of the slip velocity rather than the slip length (see \cite{premlata2019basset} for a recent work).
It is worth-noting that the slip-length tends to infinity when reaching full-slip conditions while the slip-velocity remains finite.
In addition, the present model allows for framing results beyond the full-slip case which could prove relevant for mechanism such as the Leidenfrost effect\cite{vakarelski2011drag} or air injection \cite{du2017maintenance}.

The flow around a slipping sphere is solved analytically using Stokes' approach based on a streamfunction formulation. The two components of the velocity are defined as:
\begin{equation}
 u_r=\frac{1}{r^2\mathrm{sin} \theta}\frac{\partial\psi}{\partial\theta}, \quad u_\theta=-\frac{1}{r\ \mathrm{sin}\theta}\frac{\partial\psi}{\partial r},  
\end{equation}
where the polar velocity $u_\varphi$ is assumed to be null. This streamfunction is solution of the bi-harmonic equation:
\begin{equation}
\nabla^2\left(\nabla^2\psi\right)=0,
\end{equation}
where the Laplacian $\nabla^2$ in spherical coordinates writes: 
$$
\nabla^2=\frac{\partial^2}{\partial r^2}+\frac{1}{r^2 \sin{ \theta}}\frac{\partial}{\partial\theta}\left[\sin{\theta}\frac{\partial}{\partial\theta}\right].
$$\\

The boundary conditions in the present problem are:
\begin{itemize}
    \item[-] A no-penetration condition at $r=R$:
\begin{equation}
u_r\mathrm{=0}\Rightarrow \frac{1}{r^2\mathrm{sin}  \theta}\frac{\partial\psi}{\partial\theta}= 0\ \ \forall\ \theta.
\end{equation}
    
    \item[-] A slip condition at $r=R$, that is $u_\theta(r=R,\theta)=U_{slip}f(\theta)$ where $U_{slip}$ is the slip velocity and $f(\theta)=\exp(im\theta)$ is chosen as the azimuthal harmonic $\theta$ (spherical harmonic). Based on the bi-harmonic equation for the stream function, $f(\theta)$ only admits a restricted number of solutions, that is $\nabla^2 f(\theta)= \omega f(\theta)$.
The solution that induces the least amount of friction (i.e. the largest $\omega$) is the first eigenvalue $m=1$ which also provides the least amount of dissipation/the smallest drag coefficient. Note that $m=0$ recovers the no-slip condition. In that case, 
the tangential velocity on the wall of the sphere writes:

\begin{equation}
    u_\theta\mathrm{=}{-U}_{slip}\sin{\theta}\Rightarrow-\frac{1}{r\ \mathrm{sin}\theta}\frac{\partial\psi}{\partial r}={-U}_{slip}\sin{\theta}\ \ \forall\ \theta,
\end{equation}
    
    \item[-] A far-field condition $r\rightarrow\infty$:
\begin{equation}
    \psi=\frac{U_\infty r^2}{2}{\sin}^2{\theta}
\end{equation}
where $U_\infty$ and ${U}_{slip}$ denote the free-stream and slip velocities, respectively .\\
\end{itemize}

The general solution of the stream function when considering a slip velocity condition therefore writes:
\begin{equation}
    \psi\left(\mathrm{r,\theta}\right)=\frac{U_\infty r^2}{2}{\sin}^2{\theta}\left[1-\frac{3}{2}\left(\frac{R}{r}\right)\left(\frac{3U_\infty-2U_{slip}}{3U_\infty}\right)+\frac{1}{2}\left(\frac{R}{r}\right)^3\left(\frac{U_\infty-2U_{slip}}{U_\infty}\right)\right].
\label{eq:psi}
\end{equation}

The derivation of Eq.~(\ref{eq:psi}) is presented in details in the supplementary material. \\

The radial and tangential velocity therefore write:
\begin{equation}
 u_r\left(\mathrm{r,\theta}\right)=\frac{U_\infty}{2}\mathrm{cos\theta}\left[2-3\left(\frac{R}{r}\right)\left(\frac{3U_\infty-2U_{slip}}{3U_\infty}\right)+\left(\frac{R}{r}\right)^3\left(\frac{U_\infty-2U_{slip}}{U_\infty}\right)\right]   ,
\end{equation}
\begin{equation}
 u_\theta\left(\mathrm{r,\theta}\right)=\frac{U_\infty}{4}\sin{\mathrm{\theta}}\left[3\left(\frac{R}{r}\right)\left(\frac{3U_\infty-2U_{slip}}{3U_\infty}\right)+\left(\frac{R}{r}\right)^3\left(\frac{U_\infty-2U_{slip}}{U_\infty}\right)-4\right].
\end{equation}
Consequently, the total drag force applied on a SH sphere can be formulated in terms of the slip velocity as follows:
\begin{equation}
F_{Total}^{SH}=6\pi\mu_lRU_\infty\left[1-\frac{2}{3}\frac{U_{\mathrm{slip}}}{U_\infty}\right],
\label{eq:Drag Force}
\end{equation}
which leads to the following expression for the drag coefficient:
\begin{equation}
    C_{D}^{SH}=\frac{24}{Re}\left[1-\frac{2}{3}\frac{U_{\mathrm{slip}}}{U_\infty}\right].
    \label{eq:drag expression slip}
\end{equation}

For free-slip surface, the tangential shear stress is absent, hence,
the liquid slips with a velocity $U_{slip}=U_\infty/2$. As a result, the theoretical drag would be:

\begin{equation}
      C_{D}^{SH}=\frac{16}{Re}.
      \label{eq:slip drag}
\end{equation}

The development of the theoretical model and its validation using numerical simulations are presented in details in the supplementary material.\\

For realistic surfaces, where roughness elements exist, the slip velocity can hardly reach the full slip case given in (\ref{eq:slip drag}) because the plastron may not entirely cover the surface of the sphere. Henceforth, $U_{slip}$ has to be expressed in terms of the characteristics of the surface which is assumed here as a porous medium. 
In what follows, the slip velocity is determined as a function of the geometric features of a SH-surface where we consider the flow inside the textured surface. The expression should involve two parameters commonly used in porous media: $(i)$ the porosity and $(ii)$ the tortuosity of the surface. The aim is to determine the mean slip velocity $U_{slip}$ and determine whether it can be approximated as scaling laws. 
We propose an analogy with the flow across micro channels where the tortuosity and the porosity can be controlled independently. The resulting homogenised slip is subsequently used in Eq.~\ref{eq:drag expression slip} to estimate drag.


\subsection{Homogenised boundary condition}\label{subsec:homogenised}

\begin{figure}[t!]

\includegraphics[width = 0.9 \columnwidth]{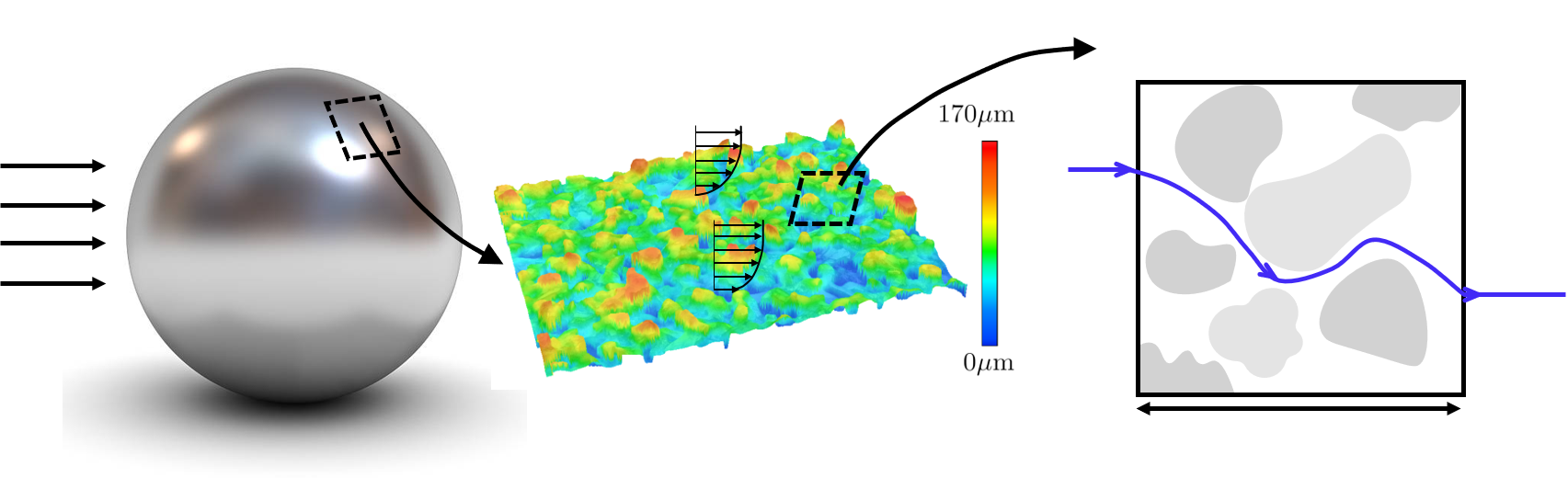}
\put(-380,150){(a)}
\put(-485,75){$U_\infty$}
\put(-420,-5){SH-Sphere in free stream}
\put(-260,150){(b)}
\put(-270,-5){Surface texture}
\put(-270,70){Slip}
\put(-290,100){No-Slip}
\put(-140,-5){Air motion inside the CV}
\put(-80,150){(c)}
\put(-145,95){{\textcolor{blue}{{$L_{\Theta}$}}}}
\put(-80,10){{{$L$}}}
\put(-80,100){Air}
\put(-65,45){Solid}
\put(-25,60){{\textcolor{blue}{{Air flow path}}}}

\caption{Illustration of the air flow through roughness elements of a (a) SH sphere. (b) Confocal microscopy analysis of a portion ($1.5 \times 1.0$ mm$^2$) from a flat plate covered by a fine grain SH coating. The colorscale (blue-to-red) indicates the surface roughness height $(0 - 170\mu m)$. Notice the random spatial distribution of the powder particles. (c) Illustration of the gas pathline through a sample of random roughness elements.}
\label{fig:Air motion}
\end{figure}

As depicted in Fig.~\ref{fig:Air motion}, the interfacial boundary condition alternates between no-slip and slip because of the random positioning of the roughness elements. As a result, the mean slip velocity varies under the effect of porosity. Taking a closer look to the movement of air between the roughness elements shows the tortuous path the air may follow which can also alter slip at the wall.
The length of the path is referred to as the tortuosity $\Theta$ of the Control Volume (CV). It is defined as the ratio of the actual path followed by the air (${L_{\Theta}}$) and length of the control volume ($L$). The arrangement of roughness elements from Fig.~\ref{fig:Air motion} yields a tortuosity ${{\Theta}>1}$. Whereas, it is anticipated that if the surface is composed of structured elements, pillars for instance, air will experience less resistance. Thereby, flow paths will be relatively straight, ${L_{\Theta}}\approx1$.

\begin{figure}[t]

\includegraphics[width = 0.9 \columnwidth]{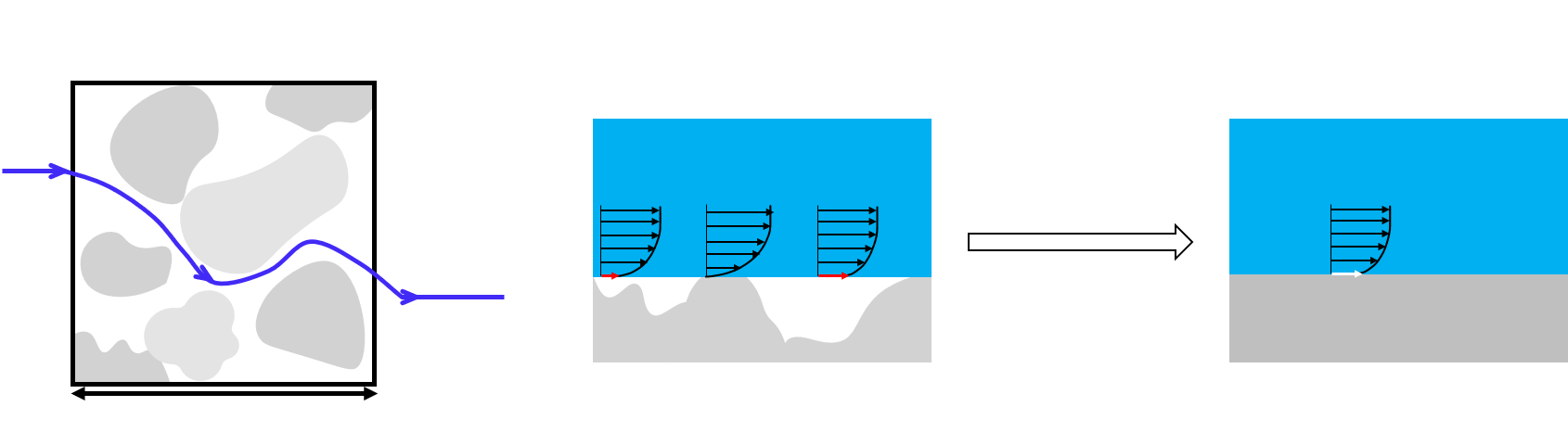}
\put(-458,80){{\textcolor{blue}{{$L_{\Theta}$}}}}
\put(-395,80){Air}
\put(-380,28){Solid}
\put(-398,2){{{$L$}}}
\put(-450,-8){Air motion inside the plastron}
\put(-342,45){\textcolor{blue}{Air flow path}}
\put(-265,-8){Local slippage}
\put(-285,85){Liquid}
\put(-285,70){\textcolor{red}{{Slip}}}
\put(-220,70){\textcolor{red}{{Slip}}}
\put(-258,70){No-Slip}
\put(-225,28){Air}
\put(-285,28){Solid}
\put(-98,85){Liquid}
\put(-98,28){Solid}
\put(-98,70){\textcolor{white}{{Slip}}}
\put(-70,70){$U_\infty$}
\put(-70,35){\textcolor{white}{$U_{slip average}$}}
\put(-152,62){BC}
\put(-172,42){homogenisation}

\caption{Homogenisation of the boundary condition at the air/liquid interface in terms of the geometric characteristics of the SH surface}
\label{fig:homogenisation}
\end{figure}
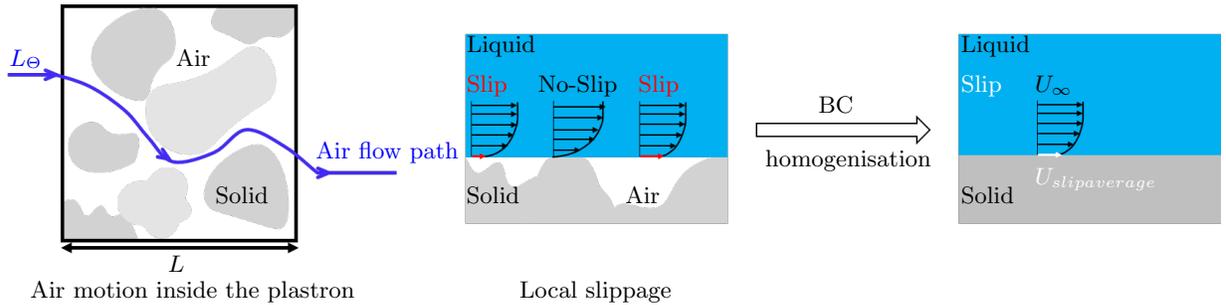

Fig.~\ref{fig:homogenisation} further illustrates how the local slip velocity varies for the case of randomly distributed roughness elements. More precisely, slip can be highly intermittent since the air void is different from one air pocket to another. The homogenisation approach hence consists in finding an expression for the mean slip velocity that accounts for the geometric characteristics of the plastron and the motion of air inside it. To homogenise the BC on the composite interface, numerical investigations were carried out. The study was conducted based on the classical approach of a CV. Details regarding the numerical simulations are provided in the supplementary material.

\begin{figure}[t]

{\includegraphics[width = .7 \columnwidth]{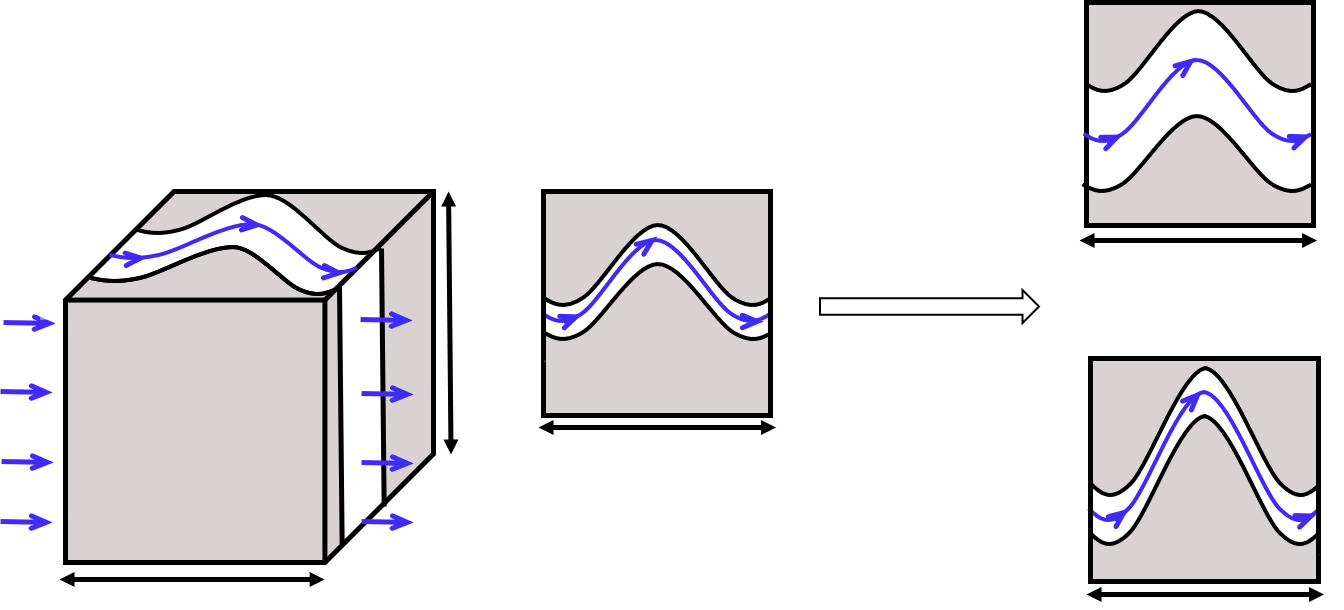}}
\put(-230,80){{\textcolor{blue}{{$L_{\Theta}$}}}}
\put(-85,130){{\textcolor{blue}{{$L_{\Theta}$}}}}
\put(-85,30){{\textcolor{blue}{{$L_{\Theta}$}}}}
\put(-185,38){{L}}
\put(-270,-5){{Air motion inside the plastron}}
\put(-193,60){{Solid}}
\put(-197,118){{$\Phi_1, \Theta_1$}}
\put(-140,85){{Variation of }}
\put(-132,70){{$\Phi$ and $\Theta$}}
\put(-308,-3){{L}}
\put(-315,42){{Solid}}
\put(-400,45){\textcolor{blue}{{Air inflow}}}
\put(-40,-5){{L}}
\put(-47,17){{Solid}}
\put(-70,75){{$\Phi=\Phi_1,\Theta> \Theta_1$}}
\put(-40,89){{L}}
\put(-47,111){{Solid}}
\put(-70,169){{$\Phi>\Phi_1,\Theta= \Theta_1$}}
\caption{Schematic of a porous medium model over a Control Volume CV (width $L$ and height $h$) inside the air layer on the surface of a SH sphere. Black regions are occupied by the solid while white regions are occupied by the air.}
\label{fig:var T Phi}
\end{figure}
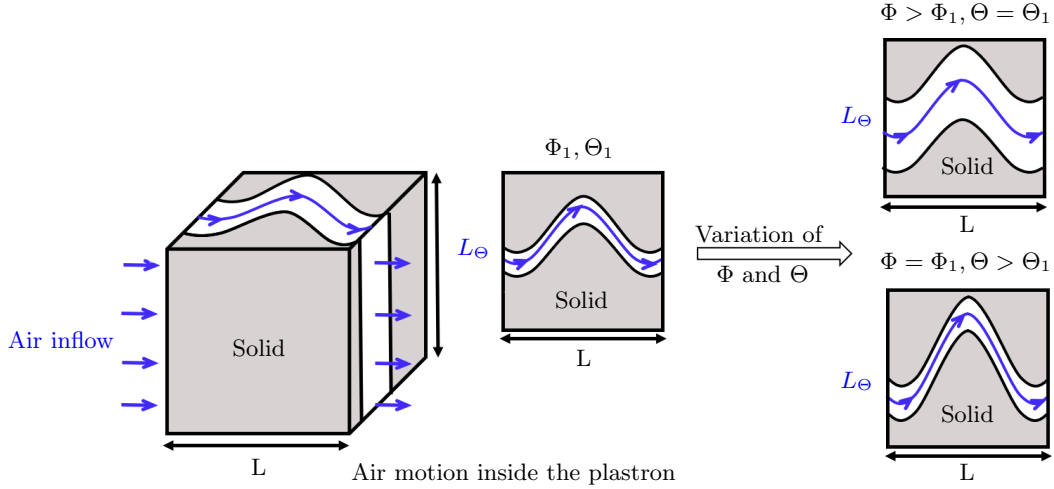

To explore the impact of surface texture, the slip velocity was computed numerically for different cases where the porosity and the tortuosity of the domain were varied independently. As presented in Fig.~\ref{fig:var T Phi}, air is forced to follow a sinusoidal path that is longer than the length of the CV. 
The porosity $\Phi$, which corresponds to the fraction of air occupied in the CV is defined as:
\begin{equation}
 \mathrm{\Phi}=\frac{A_g}{A_g+A_s},
\end{equation}
where $A_g$ and $A_s$ are the gas and solid volumes, respectively. The air flow path was computed using the geometric definition of the tortuosity; In the case of sinuous channels, ${L_{\Theta}}$ is equal to the length of the channel walls. Fig.~\ref{fig:var T Phi} further indicates how $\Phi$ and $\Theta$ are varied independently. Considering a first configuration where the porosity and the tortuosity are $\Phi_1$ and $\Theta_1$, respectively. The porosity of the CV is increased ($\Phi>\Phi_1$), at a constant $\Theta$, by increasing the distance between the walls of the channel. Increasing the tortuosity ($\Theta>\Theta_1$) was achieved by increasing the amplitude of the sinuous shape of the channel. \\

The fluid in the CV is subjected to both pressure and shear forces from the outer fluid. To account for both effects, a linear combination between a Shear Driven Flow (SDF) and Pressure Driven Flow (PDF) setup is proposed.  For each type of flow, the air motion within the channels was investigated for two different scenarios. In each case, one of the two geometric characteristics  was varied while the other was kept constant. 
For instance, given a defined geometric tortuosity $\Theta_1$, simulations were performed for different values of the porosity $\Phi$. Note that the finite CV size imposed a limited range for varying both $\Phi$ and $\Theta$. The chosen values for $\Theta$ ranged between $\Theta=1.024$ and $\Theta=1.463$ whereas the porosity $\Phi$ in the CV varied within $5\%-45\%$. For each simulation, the normalised slip velocity $U_{\mathrm{slip}}^{\star}={U_{\mathrm{slip}}}/{{U_\infty}}$ is calculated as the spatial mean value of the velocity on the upper surface of the CV. \\

\begin{figure} [t]
\begin{minipage}{0.48\linewidth}
\centering
{\label{fig:PressureSlipVsP}
\includegraphics[width=1\textwidth]{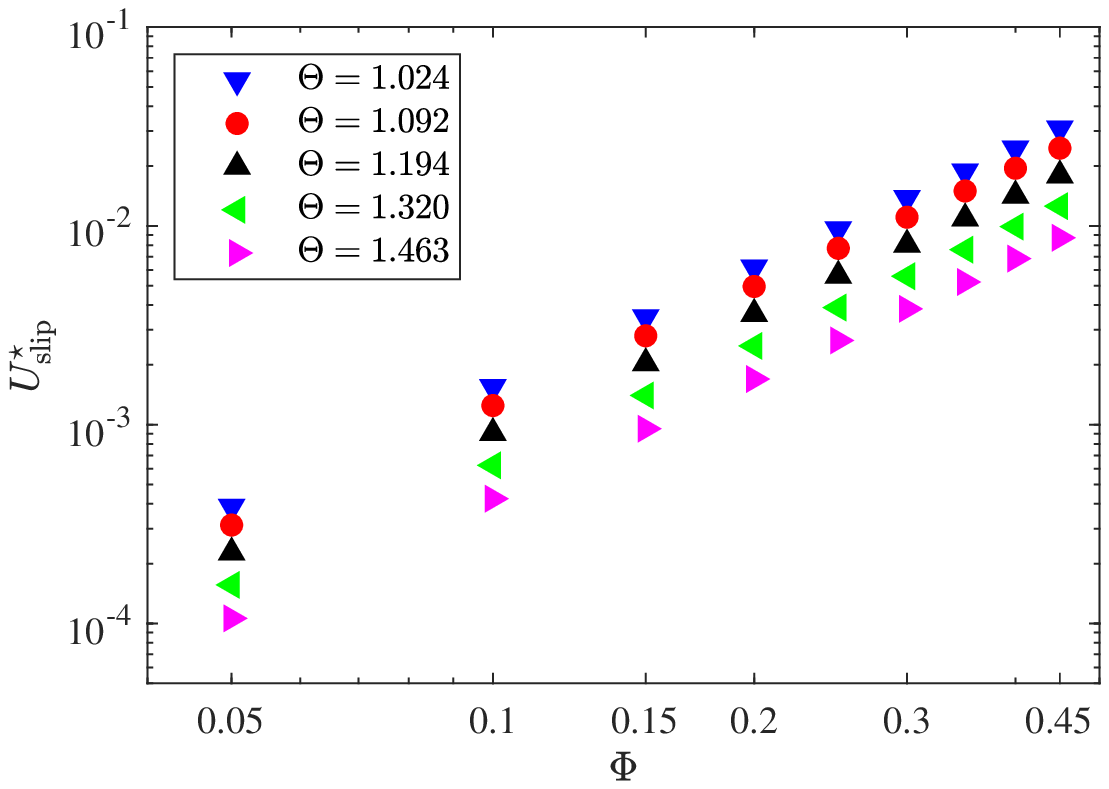}}
\subcaption{Constant tortuosity, $U_{\mathrm{slip}}^{\star}={U_{\mathrm{slip}}}/{{U_\infty}}$ }
\end{minipage}
\begin{minipage}{.48\linewidth}
\centering
\includegraphics[width=1\textwidth]{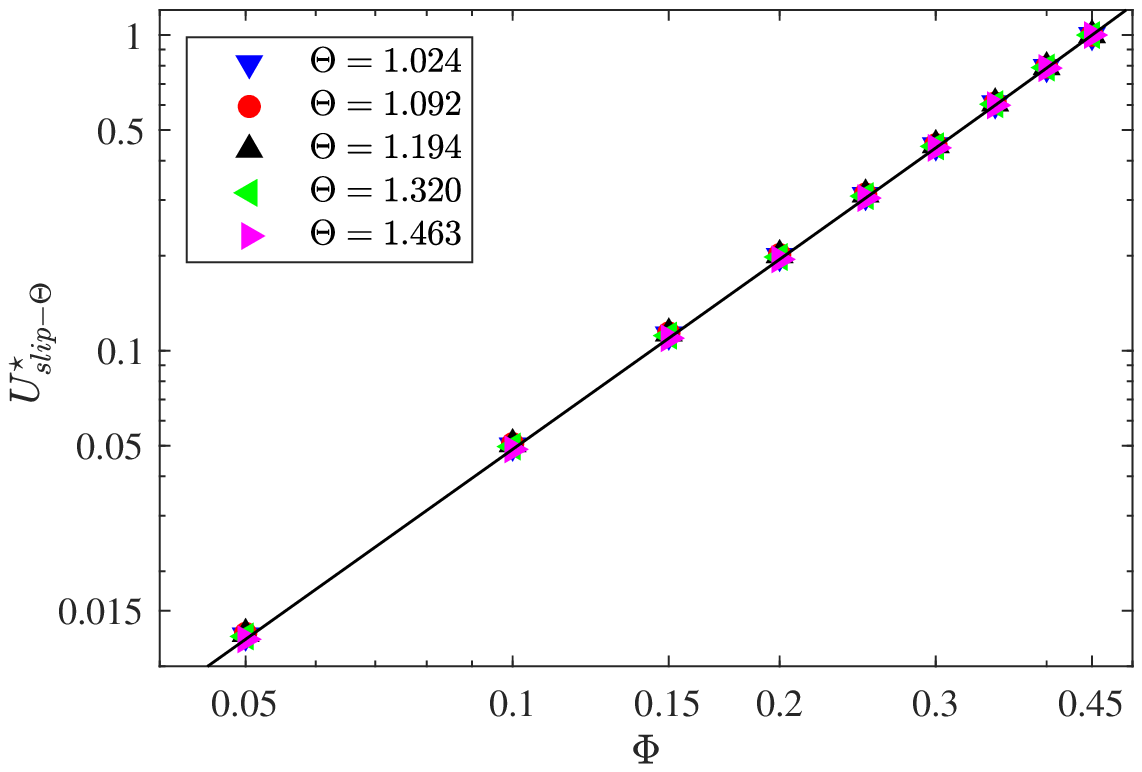}

{\label{fig:PressureSlipVsP_Norm}}
\subcaption{Scaled velocity, $U_{slip-\Theta}^{\star}=U_{\mathrm{slip}}^{\star}/{U_{\mathrm{slip}}^{\star}(\Phi=0.45)}$}
\end{minipage}

\begin{minipage}{0.48\linewidth}
\centering
\includegraphics[width=1\textwidth]{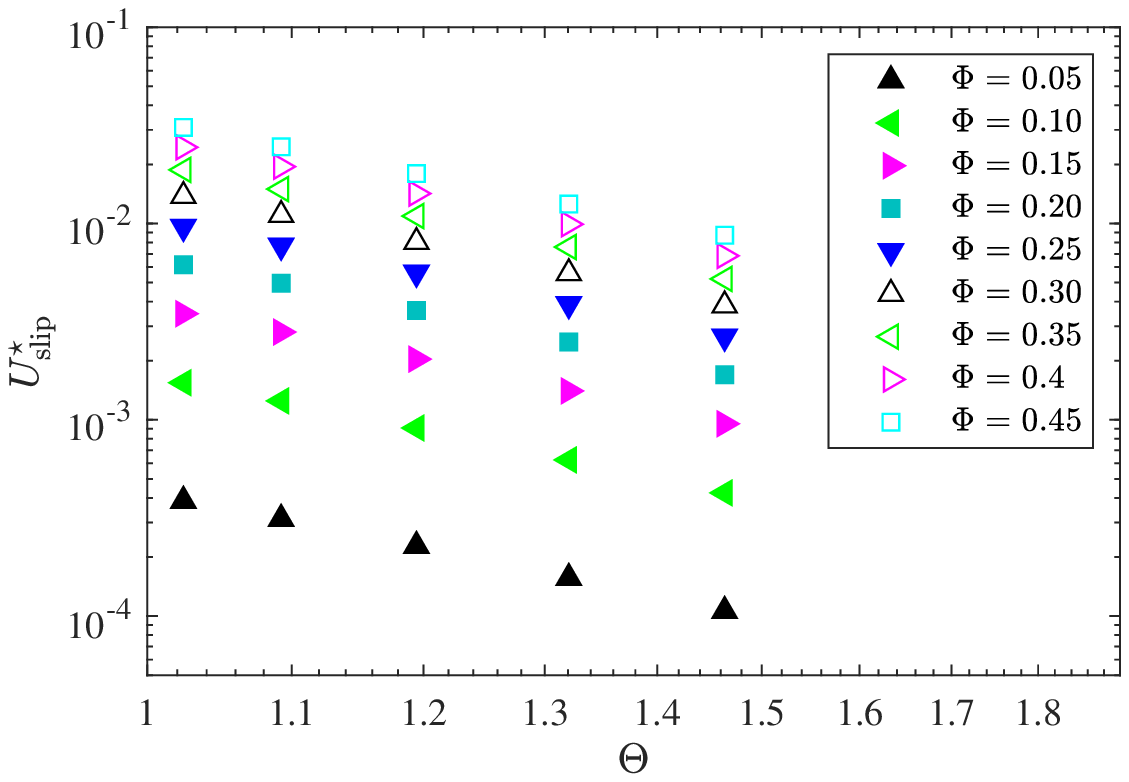}
\subcaption{Constant Porosity}
\end{minipage}
\begin{minipage}{.48\linewidth}
\centering
\includegraphics[width=1\textwidth]{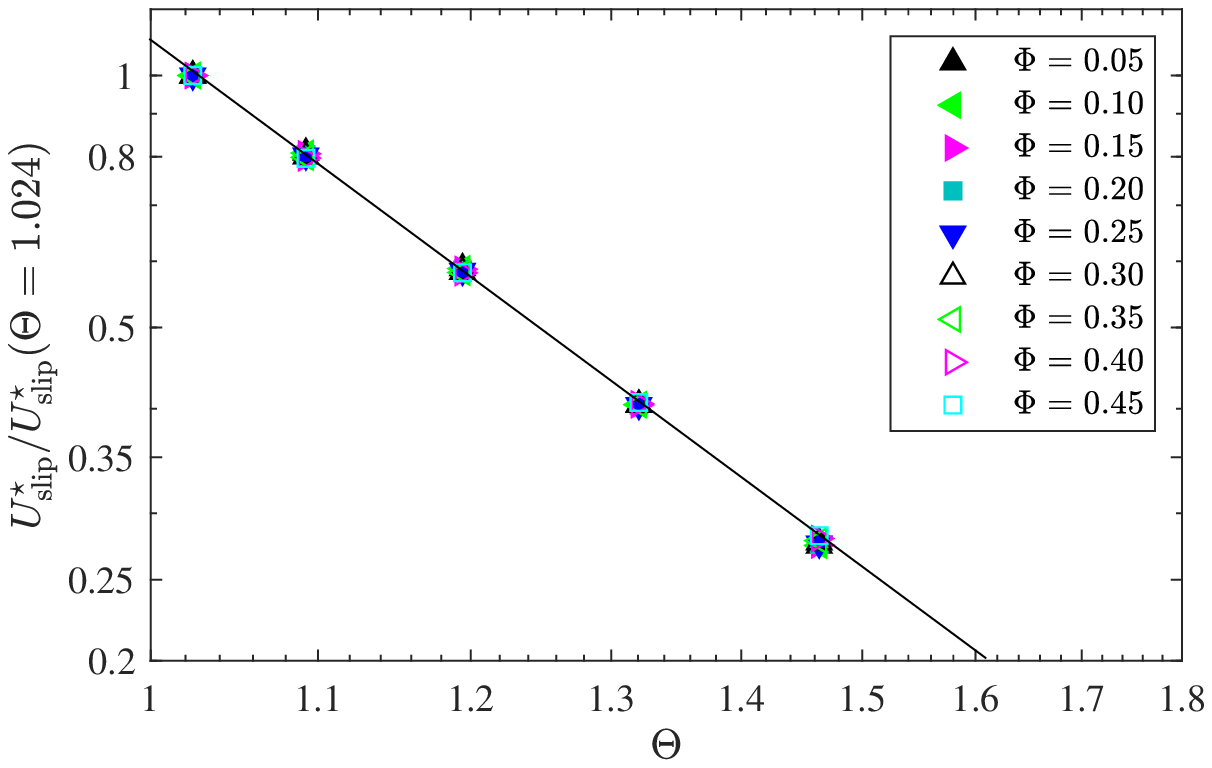}
\subcaption{Scaled velocity, $U_{slip-\Phi}^{\star}=U_{\mathrm{slip}}^{\star}/{U_{\mathrm{slip}}^{\star}(\Theta=1.024)}$}
\end{minipage}

{\caption{Slip velocity variation in terms of the porosity and tortuosity of the domain: Pressure driven flow case.}
\label{fig:PowerLaws1}}

\end{figure}

For the PDF, the variation of $U_{\mathrm{slip}}^{\star}$ with respect to the geometric characteristics of the CV is shown in Fig.~\ref{fig:PowerLaws1}. At a constant value of $\Theta$, Fig.~\ref{fig:PowerLaws1}(a) shows that the normalised slip velocity increases with increasing $\Phi$. This is expected since a larger porosity provides a larger slip area between the liquid and the gas. However, a deeper look to Fig.~\ref{fig:PowerLaws1}(a) reveals that the increase in tortuosity results in a remarkable decrease of the slip velocity. In fact, for $L_{\Theta}$ less than $1.2 L$, the slip velocity is approximately $40\%$ less compared to the smallest values of the investigated tortuosity. 
Fig.~\ref{fig:PowerLaws1}(c) shows the effect of the tortuosity on the mean slip velocity for constant values of $\Theta$. A similar behaviour is observed and slip is favoured when a higher percentage of void and lower values of $L_{\Theta}$ are considered. Fig.~\ref{fig:PowerLaws1}(b,d) illustrate the variation of the scaled slip velocity defined as  $U_{slip-\Theta}^{\star}=U_{\mathrm{slip}}^{\star}/{U_{\mathrm{slip}}^{\star}(\Phi=1.024)}$ and $ U_{slip-\Phi}^{\star}=U_{\mathrm{slip}}^{\star}/{U_{\mathrm{slip}}^{\star}(\Theta=1.024)}$ for constant tortuosity and constant porosity simulations, respectively. 
From Fig.~\ref{fig:PowerLaws1}(d), more than $70\%$ of the slip velocity is lost for all cases, when $\Theta < 1.5L$ and for different values of $\Phi$, which results in a remarkable decrease performances.\\

\begin{figure} [h]
\begin{minipage}{0.48\linewidth}
\centering
{\label{fig:PressureSlipVsP1}
\includegraphics[width=1\textwidth]{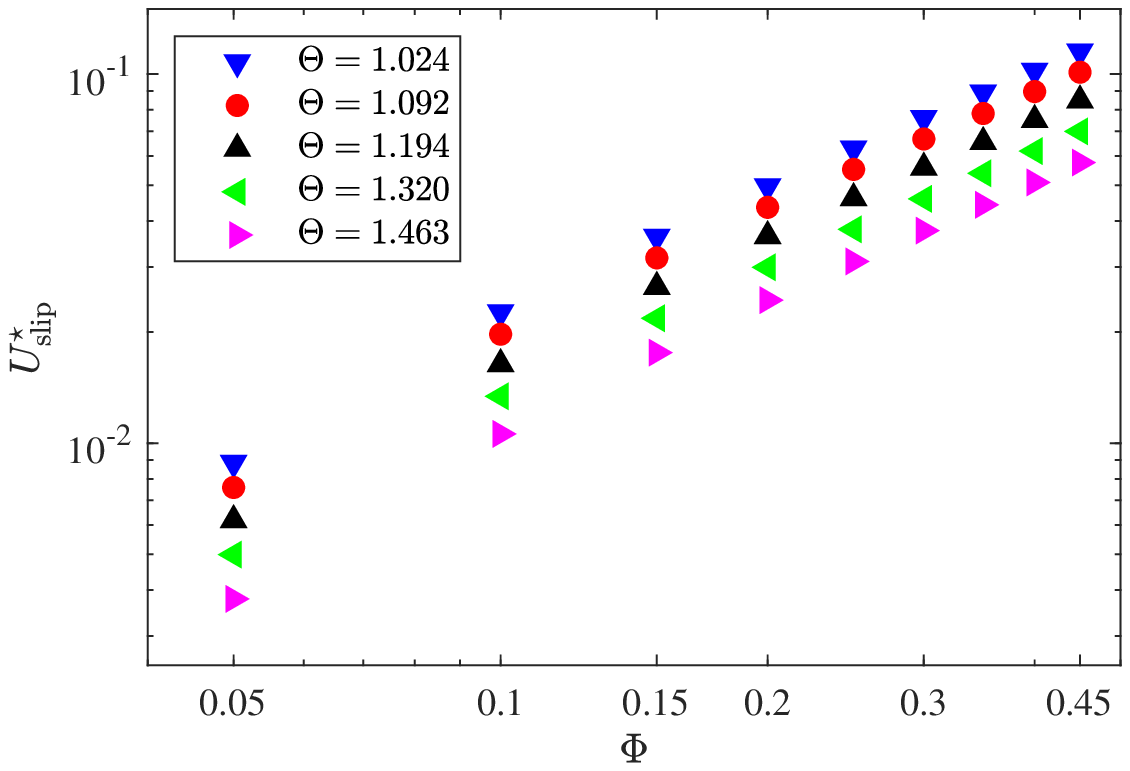}}
\subcaption{Constant tortuosity, $U_{\mathrm{slip}}^{\star}={U_{\mathrm{slip}}}/{{U_\infty}}$}
\end{minipage}
\begin{minipage}{.48\linewidth}
\centering
\includegraphics[width=1\textwidth]{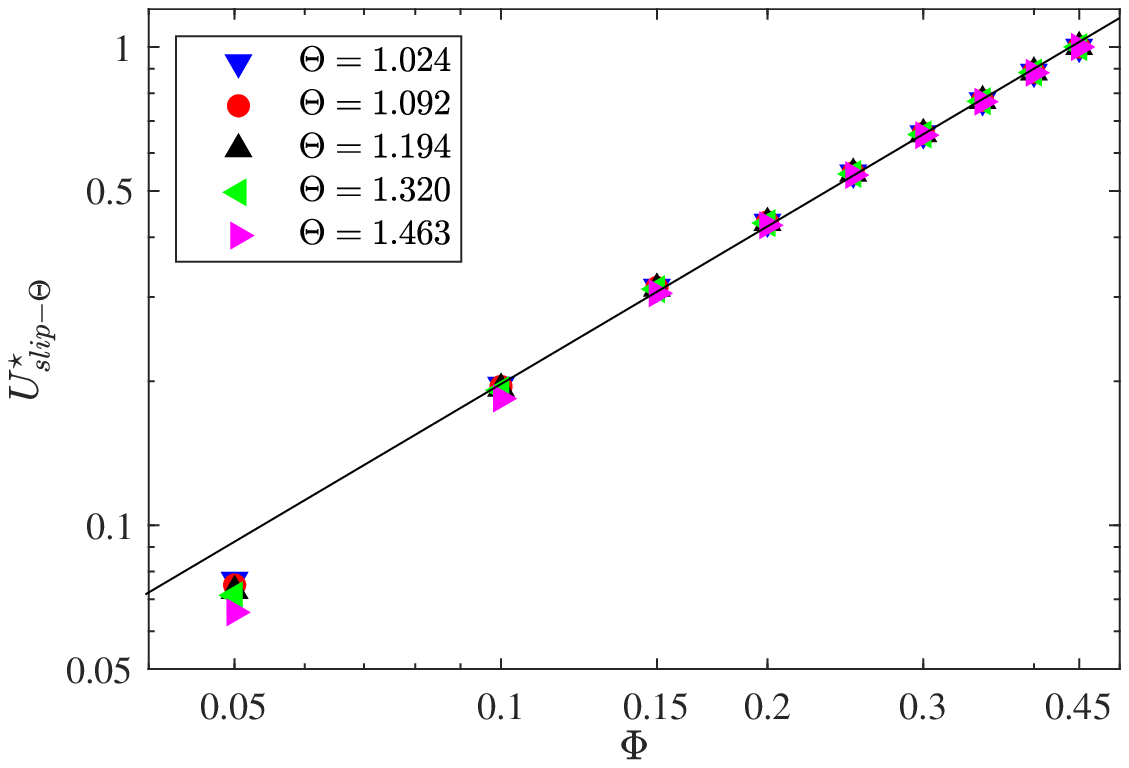}

\subcaption{Scaled velocity, $U_{slip-\Theta}^{\star}=U_{\mathrm{slip}}^{\star}/{U_{\mathrm{slip}}^{\star}(\Phi=0.45)}$}
\end{minipage}

\begin{minipage}{0.48\linewidth}
\centering
\includegraphics[width=1\textwidth]{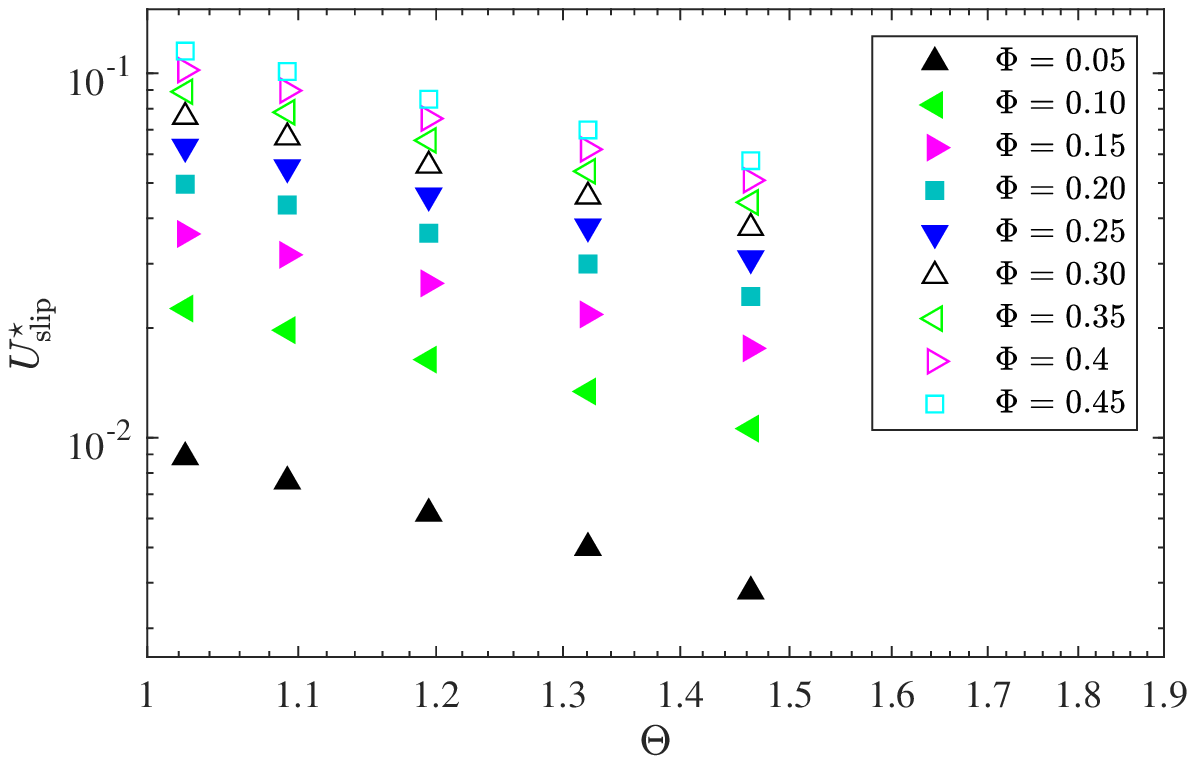}
\subcaption{Constant Porosity, $U_{\mathrm{slip}}^{\star}={U_{\mathrm{slip}}}/{{U_\infty}}$}
\end{minipage}
\begin{minipage}{.48\linewidth}
\centering
\includegraphics[width=1\textwidth]{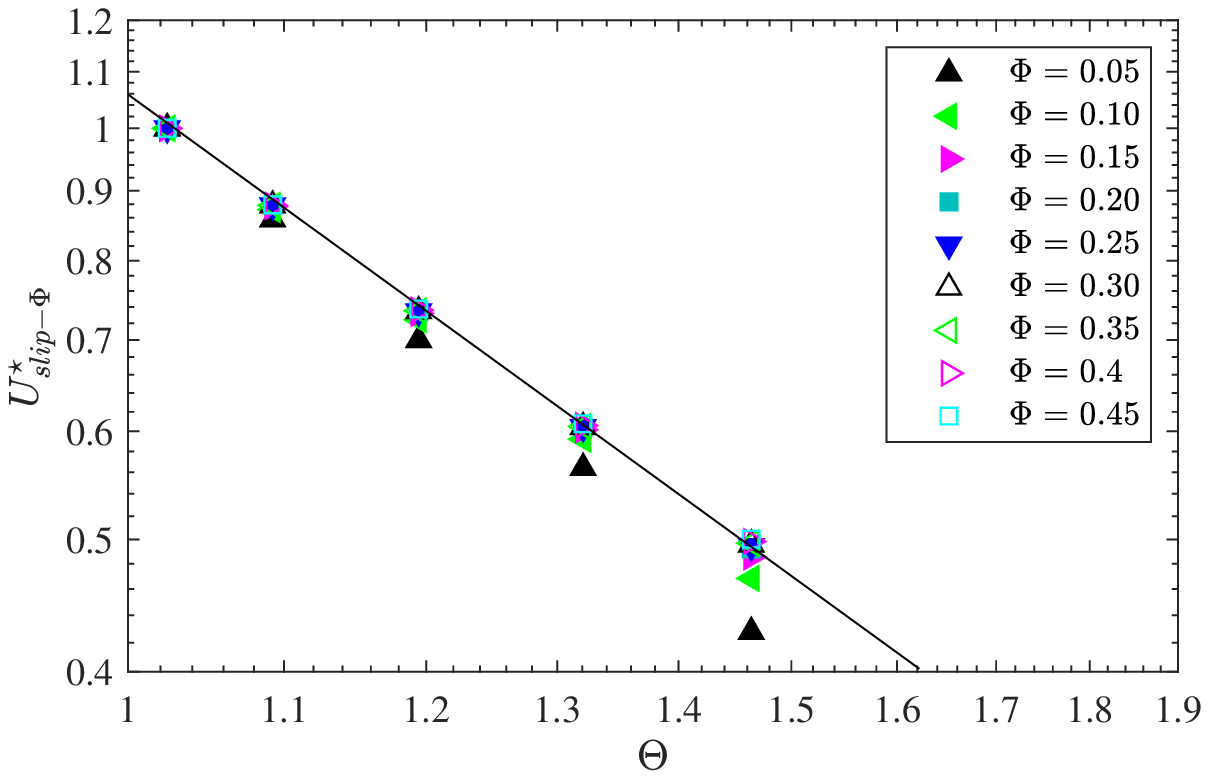}
\subcaption{Scaled velocity, $U_{slip-\Phi}^{\star}=U_{\mathrm{slip}}^{\star}/{U_{\mathrm{slip}}^{\star}(\Theta=1.024)}$}
\end{minipage}

{\caption{Slip velocity variation in terms of the porosity and tortuosity of the domain: Shear driven flow case }
\label{fig:PowerLaws2}}
\end{figure}

For the SDF, variations of the normalised slip velocity in terms of the geometric characteristics is similar to that obtained for the PDF. Nonetheless, larger slip velocities are obtained for SDF in comparison to the PDF case. Tab.~\ref{tab:comparison Shear-pressure} indicates that the slip velocity ratio, defined as ${U_{slip}^{\star}(SDF)}/{U_{slip}^{\star}(PDF)}$ essentially varies linearly with both the tortuosity and the porosity. The comparison is performed for different values of tortuosity and at lowest and highest values of $\Phi$. A factor of $35$ between the slip velocity for the two flow scenarios is obtained at $\Phi=0.05$ and $\Theta=1.463$. These findings emphasise the important contribution of shear to the global slip velocity. This contribution is at its highest for the lowest values of void percentage. 

\begin{table}[t!]
\begin{ruledtabular}
\begin{tabular}{cccccc}
& $\Theta=1.024$ & $\Theta=1.092$ & $\Theta=1.194$&$\Theta=1.320$&$\Theta=1.463$ \\ 
$\Phi=0.05$  & $22.8752$  & $24.2552$ & $27.1888$ &$31.9064$&$35.6491$\\ 

$\Phi=0.45$  & $3.7263$  & $4.1133$ & $4.7266$ &$5.5651$&$6.6086$\\ 
\end{tabular}
\end{ruledtabular}

\caption{Slip-velocity ratio ${U_{slip}^{\star}(SDF)}/{U_{slip}^{\star}(PDF)}$  between the SDF case and the PDF case at constant tortuosity.}
\label{tab:comparison Shear-pressure}
\end{table}

From Fig.~\ref{fig:PowerLaws1}(b,d) and Fig.~\ref{fig:PowerLaws2}(b,d), it is clear that the slip velocity scales as power laws with the tortuosity $\Theta$ and the porosity $\Phi$ of the domain such that:
\begin{equation}
U_{slip-}^{\star} \approx C {\Theta ^a} {\Phi ^b},
\end{equation}
where $C$, $a$, and $b$ are constants calculated form the numerical results.

The values of the constants for the two scenarios are reported in Tab.~\ref{tab:Constant power Law}. Using the linear properties of Stokes' flow for both the shear driven flow and the pressure driven flow, the expression of the total normalised slip velocity becomes:
\begin{equation}
    {U_{slip}^{\star}} = {U_{slip-SDF}^{\star}} + {U_{slip-PDF}^{\star}}
\end{equation}

\begin{table}[t]
\begin{ruledtabular}
\begin{tabular}{cccc}
& C & a & b \\ 
Pressure Driven Flow  & $0.15$  & $-3.6$ & $2$ \\ 
Shear Driven Flow     & $0.31$  & $-2$ & $1.2$  \\ 

\end{tabular}
\end{ruledtabular}

\caption{Power law constants obtained from numerical simulations}
\label{tab:Constant power Law}
\end{table}


The obtained expression can be directly applied to obtain an average value of the slip velocity given that both $\Phi$ and $\Theta$ are known. Once calculated, $U_{slip}^{\star}={U_{\mathrm{slip}}}/{{U_\infty}}$ can be implemented in the theoretical expression of the drag coefficient (Eq.\ref{eq:drag expression slip}) to predict drag on a SH-sphere with known porosity and tortuosity.\\

In what follows, the drag of falling SH-spheres at low Reynolds numbers is investigated experimentally. The aptitude towards drag reduction of SH-spheres with randomly distributed roughness elements is first investigated. A comparison between the predicted drag on these spheres using the proposed model and the experiments follows. An additional sphere with carefully designed SH-texture is later proposed with the aim to lower $\Theta$ while keeping $\Phi$ within the same range.

\section{Experimental setup}
\begin{figure}[t!]
\includegraphics[width = 0.55 \columnwidth]{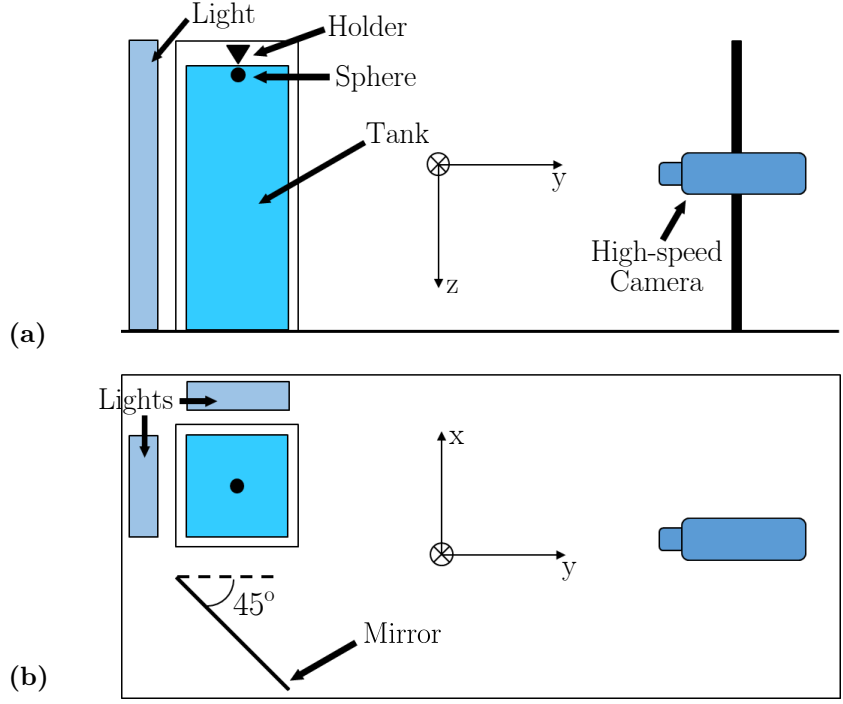}
\put(-315,135){\textbf{\textcolor{black}{\textbf{\normalsize(a)}}}}\put(-315,5){\textbf{\normalsize\textcolor{black}{(b)}}}
\caption{ The falling sphere experimental set-up: (a) side view, (b) top view.}
\label{setup}
\end{figure}

Apart from the working fluid, the experimental setup used in this study is identical to the one described in detail in Castagna {\it et al.}\cite{Castagna2018,Castagna2021}.
Falling sphere experiments were performed in a transparent tank ($100\times100$  $mm^{2}$ square cross-section and 650 mm height) filled with glycerine (see Fig.~\ref{setup}). 

All tests were performed at room temperature equal to $20\pm1$ $^{\circ}C$, which was verified by a thermocouple dipped in the glycerine.
The glycerine density $\rho_{l}$ and dynamic viscosity $\mu_{l}$ were evaluated using the empirical formula proposed by Cheng\cite{Cheng2008}.
The value $\mu_{l}=1.41$ Pa$\cdot$s valid for pure glycerine at 20 $^{\circ}C$ was further experimentally verified with a Fungilab{\texttrademark} viscometer. At the beginning of the test, an electromagnetic holder was used to gently dip the spheres below the glycerine surface in order to assure a null velocity at release time. Stainless steel spheres with nominal diameters $d$ equal to $5$, $8$ and $10$ mm were taken as reference spheres. The trajectory of the falling sphere was recorded by a Phantom V341 high-speed camera at a 2560$\times$1100 px$^{2}$ resolution, resulting into a conversion factor of 0.3 mm px$^{-1}$.

\begin{figure}[t!]
\includegraphics[width = .62 \columnwidth]{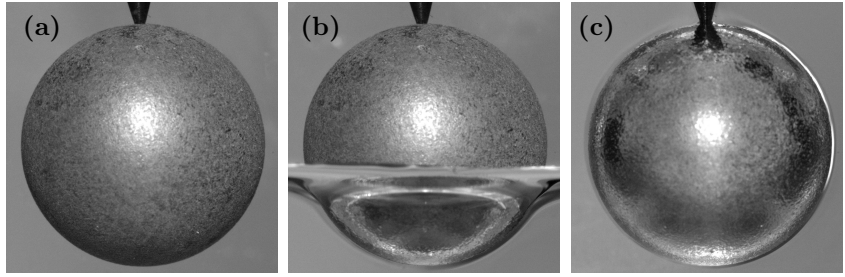}
\put(-310,90){\textbf{\textcolor{black}{\textbf{\normalsize(a)}}}}\put(-205,90){\textbf{\normalsize\textcolor{black}{(b)}}}
\put(-100,90){\textbf{\normalsize\textcolor{black}{(c)}}}
\caption{
Visual evidence of the air layer around a super-hydrophobic sphere (SH-1 coating, $d=15$ mm diameter) progressively dipped in glycerine. (a), air. (b), partially dipped. (c), fully immersed. Image (b) underlines the difference between the area above the glycerine surface and the part of the sphere dipped in the glycerine. The bright appearance in images (b) and (c) testifies the establishment of the air layer. The thickness of the air layer is qualitatively discernible in both images (b) and (c).
}
\label{fig:2Suppl}
\end{figure}

The recording frame rate was adapted to the sphere falling velocity, ranging from $200$ fps for the $d=5$ mm up to 800 fps for the $d=10$ mm spheres. 
An additional mirror was set at $45^o$ with respect to the tank (see Fig.~\ref{setup}(b)). The recorded images allowed for recovering the three-dimensional (3D) position of the sphere.
The post-processing of the recorded videos which enabled the reconstruction of the 3D displacement of the falling sphere was performed with the commercial software MATLAB\textsuperscript{\textregistered}. A cross-correlation code able to achieve sub-pixel accuracy by means of a Gaussian fit of the correlation peak\citep{Raffel2007} was developed. The accuracy was estimated to be approximately $\pm$ $0.06$ px, that is lower than $0.4\%$ of the smallest investigated diameter.
It was found that the sphere motion in the transverse plane $(x-y)$ is negligible, indicating the absence of lift. Therefore, in the remainder of the paper, we only focus on vertical motion, hence drag.

Terminal velocities $U_{\infty}$ were evaluated in the range $0.07-0.23$ m/s, which corresponds to a terminal Reynolds number ($Re_{\infty}=\rho_{l}U_{\infty}d/\mu_{l}$) lying within $0.3-2.5$. 
Confinement effects induced by the finite size of the test section were compensated for
following the technique proposed by Di Felice \cite{DiFelice1996}: 
%
\begin{equation}
\frac{U_{\infty}^{bou}}{U_{\infty}^{unb}}=\left(\frac{1-\delta}{1-0.33\delta}\right)^{\alpha},
\label{eq:Vbound}
\end{equation}
where the ratio of the bounded terminal velocity $U_{\infty}^{bou}$ with respect to the unbounded $U_{\infty}^{unb}$ counterpart depends on the blockage factor $\delta=d/D_{tank}^{eq}$. In the latter expression, $D_{tank}^{eq}=2W/\sqrt{\pi}$ ($W$ being the tank width) is the equivalent diameter of the tank. In this study, the blockage factor $\delta$ is lower than 0.09. Finally, the exponent $\alpha$ in Eq.~(\ref{eq:Vbound}) was evaluated by the following expression \cite{DiFelice1996}:
\begin{equation}
\frac{3.3-\alpha}{\alpha-0.85}=0.27Re_{\infty}^{0.64}. \\
\end{equation}
%
As explained in Castagna {\it et al.}\cite{Castagna2018}, SH coatings were produced by a spray method technique suitable for macroscopic applications using a commercially available  SH paint (Ultra-Ever Dry\textsuperscript{\textregistered}). 
Two different types of surface texturing were employed to evidence the effect of the spatial distribution of roughness on drag. The random distribution of the baseline SH coating (SH-1, no added powder) was enhanced by depositing an intermediate layer made of a carbon-based powder. Two different grades were used and the corresponding coatings will be indicated hereinafter SH-2 (fine grain) and SH-3 (coarse grain) in order of increasing root-mean square surface roughness $\lambda$ values.
The latter were evaluated by 3D confocal microscopy measurements over SH flat plates (see Fig.~\ref{fig:Air motion}, with values in the range $25-142$ $\mu$m (see Tab. \ref{tab:1Suppl}). 

Contact angle measurements were performed with a digital goniometer via the sessile drop technique. Glycerine drops with a volume of 6 $\mu L$ were deposited over SH horizontal flat plates providing static contact angles ($\theta_{s}$) in the range $137-156$ deg. The highest $\theta_{s}$ was obtained with the SH-1 coating, while a $\lambda$ increase determined a $\theta_{s}$ reduction, in good agreement with the findings reported in Nilsson {\it et al.}\cite{Nilsson2010} for surface roughness in the same range of our study. The same approach was followed to evaluate the roll-off ($\theta_{r}$) and the hysteresis ($\theta_{h}$) angles by tilting the SH flat plates at a controlled rate of 0.5 deg./s. The roll-off angle ($\theta_{r}$), which was considered in this study as the lowest tilt angle that caused the drop to roll-off, reached the highest value of $5.4$ deg. in the SH-3 case, whereas a $\lambda$ decrease determined a $\theta_{r}$ reduction down to $2.6$ deg. for the SH-1 coating. The hysteresis angle ($\theta_{h}$), defined as the difference between the advancing and receding angles at the time instant where the drop starts to roll-off \cite{Gao2006}, reached the highest value of $11.6$ deg. in the SH-3 case, while it decreased down to $4.9$ deg. for the SH-1 coating with the lowest $\lambda$.
The SH coatings determined a $d$ increase 
(measurement accuracy 10 $\mu$m) with respect to the reference spheres, with a maximum $+ 10\%$ in the SH-3 case. The masses of the reference spheres (measurement accuracy $0.1$ mg) were $0.51$, $2.07$ and $4.04$ g for the spheres with diameter $5$, $8$ and $10$ mm respectively. The SH coatings determined a mass increase up to $+6\%$ in the SH-3 case. Full details of the experimental set-up, SH coatings manufacturing procedure, sphere properties and post-processing techniques can be found in \cite{Castagna2018}.

\begin{table}[t]
\begin{ruledtabular}
\begin{tabular}{cccc}
 & SH-1 & SH-2 & SH-3 \\ 
$\lambda$ [$\mu m$] & 25 $\pm$ 4 & 74 $\pm$ 12 & 142 $\pm$ 23 \\ 
$\theta_{s}$ [deg.] & 156.4 $\pm$ 5.8 & 147.3 $\pm$ 7.6 & 136.7 $\pm$ 12.0 \\ 
$\theta_{r}$ [deg.] & 2.6 $\pm$ 0.8 & 2.3 $\pm$ 0.8 & 5.4 $\pm$ 2.6 \\ 
$\theta_{h}$  [deg.] & 4.9 $\pm$ 3.1 & 5.5 $\pm$ 3.9 & 11.6 $\pm$ 4.9 \\ 
$h_{r}$ [$\mu m$] & -  & 77 $\pm$ 12 & 183 $\pm$ 96 \\ 
$h_{s}$ [$\mu m$] & - & 116 $\pm$ 48 & 209 $\pm$ 99 \\ 
${\Phi}_{LA}$   & 0.59 $\pm$ 0.06 & 0.54 $\pm$ 0.06 & 0.55 $\pm$ 0.05 \\ 
\end{tabular}
\end{ruledtabular}

\caption{Properties of the manufactured SH coatings. $\lambda$, root-mean-square surface roughness. $\theta_{s}$, static contact angle. $\theta_{r}$, roll-off angle. $\theta_{h}$, hysteresis angle. $h_{r}$, horizontal size of the roughness elements, $h_{s}$, horizontal spacing of the roughness elements. ${\Phi}_{LA}$, gas fraction. The reported uncertainties represent the 95$\%$ confidence level. The missing values in the SH-1 coating are not reported since considered not reliable, due to limitation of the adopted technique.}
\label{tab:1Suppl}
\end{table}

It is worth noting that the spheres recovered after the falling experiment were dry, which confirms that the air layer is present all the way during the drop of the sphere in the tank. The glycerine-filled tank was tall enough to allow every sphere to reach its respective terminal velocity $U_{\infty}$.

As indicated previously, various mechanisms may be responsible for the loss of the performances of SH-spheres compared to their no-slip counterparts. To verify the relevance/irrelevance of these mechanisms with respect to the free-fall time, a time scale analysis is conducted. Capillary induced deformation, Marangoni stresses, and air resistance through the plastron are considered. The time characteristic scale for each mechanism is provided in details in the supplementary material and Tab.~\ref{tab:time characteristics-1} summarises the different time scales for the present experiment. It is found that, the characteristic time scales for both the Marangoni effect and interface deformation are very large in comparison with the influence of the texture on the flow. These results further confirm the assumption that air-motion through the plastron has an important effect on the slip velocity. 

 \begin{table}[t]
\begin{ruledtabular}
\begin{tabular}{cccc}
& $t_{C}^{\star}\approx \sqrt{Re_{D}Ca_{D}}$&$t_{M}^{\star}\approx \pi {Ma}^{-1}$&$t_{P}^{\star}\approx \left(\frac{\lambda}{2R} \right)^{2} Re_{D}\frac{\nu_{l}}{\nu_{a}}$\\ 
SH-1&$ \sim 10 $ & $\sim$ ${6.4}$ x $10^{2}$ & $\sim$ ${3}$ x $10^{-3}$\\ 
SH-2& $\sim 9.6$ & $\sim$ ${6.2}$ x $10^{2}$&$\sim$ ${2.4}$ x $10^{-2}$\\ 
SH-3& $\sim 9.2$ & $\sim$ ${5.9}$ x $10^{2}$&$\sim$ ${8.7}$ x $10^{-2}$\\ 
\end{tabular}
\end{ruledtabular}

\caption{Non-dimensional characteristic times for the different physical mechanisms at stake during the experiments. $Ca_{D}=\mu_{l}U_{D}/\gamma$ denotes the capillary number, $Ma$ denotes the Marangoni number, and $\nu$ denotes the kinematic viscosity.  }
\label{tab:time characteristics-1}
\end{table}

\section{Results and discussions}
\subsection{Randomly distributed roughness elements}

Drag for SH-surfaces with randomly distributed roughness elements was first investigated and compared to the theoretical Stokes drag and that obtained using the experimental correlation\cite{Clift1978}:
\begin{equation}
    C_{D-NS}=\frac{24}{Re}\left(1+0.15{Re_\infty}^{0.687}\right)
    \label{Eq:Correlation}
\end{equation}
The subscript ${-NS}$ refers to the no-slip case.
The terminal drag coefficient in the experimental was computed using the following relation for the different SH spheres:
\begin{equation}
C_{D-Exp}= 4dg\left(\zeta-1\right)/\left(3U_{\infty}^{2}\right),
\end{equation}
where $g$ is the acceleration due to gravity and $\zeta=\rho_{s}/\rho_{l}$ is the density ratio between the metal of the sphere and the glycerine, in the range $4.8-6.1$.
\begin{figure}[t]
\centerline{\includegraphics[width = 0.7 \columnwidth]{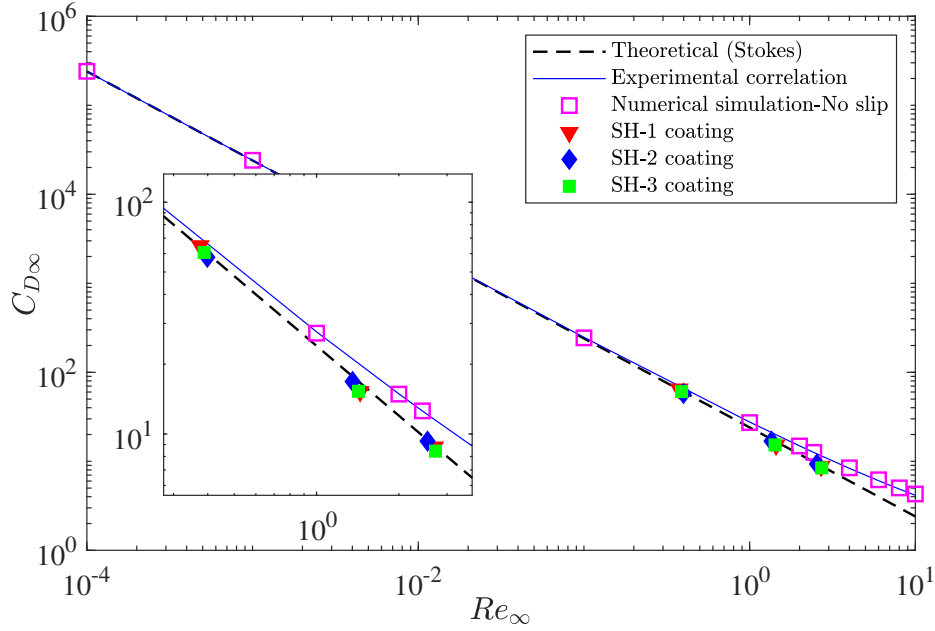}}
\caption{Terminal drag coefficient $C_{D\infty}$ as a function of the terminal Reynolds number $Re_{\infty}$.  $\textcolor{red}{\blacktriangledown}$, SH-1 coating. $\textcolor{blue}{\blacklozenge}$, SH-2 coating. $\textcolor{green}{\blacksquare}$, SH-3 coating. 
The error bars (95$\%$ confidence level) are of the same order of the size of the symbols. The dashed line shows Stokes' law ($C_{D\infty}=24/Re_{\infty}$).
}
\label{fig:CdRe}
\end{figure}
Fig.~\ref{fig:CdRe} shows the terminal drag coefficient $C_{D\infty}$ as a function of $Re_{\infty}$ for the spheres analysed in this study. The SH-coatings investigated in the current study presented a slight drag reduction in comparison to the experimental correlation. These findings emphasise the inability of SH-surfaces with randomly distributed roughness elements of producing significant drag reduction at low $Re$ flow.
For instance, at the very lowest value of $Re$, for the three different coatings, the experimentally measured drag was found to be larger than that predicted by Stokes formula ($C_{D\infty}=24/Re_{\infty}$). Nevertheless, it remains smaller than that computed using the experimental correlation.

The variation of the terminal drag coefficient as a function of $Re_{\infty}$ is further investigated
using:
\begin{equation}
   \Delta C_{D\infty} =1-\frac{C_{D-Exp}}{C_{D-NS}},
\end{equation}
All results are reported in Tab.~\ref{tab:Delta U-Free slip}.

\begin{table}[t]
\begin{ruledtabular}
\begin{tabular}{ *{6}{c} }

    \multicolumn{2}{c}{SH-1}
            & \multicolumn{2}{c}{SH-2}
                    & \multicolumn{2}{c}{SH-3}\\

  $Re_\infty$ &   $\Delta C_{D\infty}$ & $Re_\infty$  &    $\Delta C_{D\infty}$& $Re_\infty$ &   $\Delta C_{D\infty}$    \\
  
$0.375$  & 0.0570  & $0.398$ &0.1101&$0.388$ &0.0894 \\

$1.443$   &0.2325  & $1.354$ &0.1972 &$1.423$&0.2389\\

$2.702$   & 0.2376 & $2.545$ &0.2310 &$2.722$&0.2630\\

\end{tabular}
\end{ruledtabular}
\caption{Variation of the terminal drag coefficient $ \Delta C_{D\infty}$ of the SH spheres with respect to the corresponding reference sphere, as a function of the terminal Reynolds number $Re_{\infty}$.}

\label{tab:Delta U-Free slip}
\end{table}

For low $Re$ (i.e. $<0.4$), there is no significant sign of drag reduction in the case of randomly distributed SH coatings and $\Delta C_{D\infty}$ ranges from $5.7-11\%$. At $Re>1$, SH-effects become more pronounced and a drag reduction of approximately $22\%$ is reported for all experiments. As a matter of fact, the drag reduction increases with increasing value of $Re$. For instance, the SH-3 coating resulted in a drag reduction up to $26.3\%$ ($Re=2.722$). This value is still lower than predictions by analytical studies and numerical simulations \cite{McHale2011,Gruncell2013} in the case of perfect air layer or regular grooves respectively. It should be noted that no considerable effect of the surface roughness size can be derived, since the error bars (95$\%$ confidence level) of the different coatings tend to overlap. Note that the reported negligible drag reduction is always in agreement with the findings of Ahmmed {\it et al.}\cite{Ahmmed2016}. \\

\begin{figure}[t]
\centerline{\includegraphics[width = 0.3\textwidth]{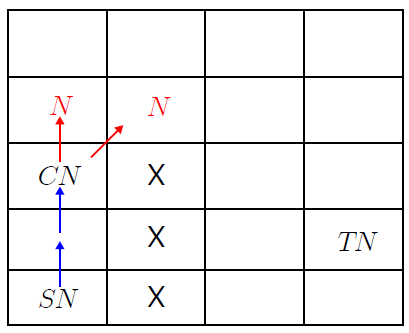}}
\caption{Schematic of the $A^{\star}$ algorithm principles. The optimal path from the starting node
$SN$ to the target node $TN$ is sought, taking into account the presence of the obstacles $X$. The next node $\textcolor{red}{N}$ is selected from the current node.}
\label{schematic algorithm}
\end{figure}

The next step aims at comparing the drag obtained experimentally for various SH-spheres with that predicted using the homogenised model. To determine the homogenised BC for the different coatings, values of the  porosity and geometric tortuosity are required. For the randomly distributed roughness spheres, the porosity for each coating type is reported in Tab.~\ref{tab:1Suppl}. To estimate the actual value of $\Theta$ for a given coating, 3D confocal microscopy images were analysed. Images were binarised setting a threshold equal to the roughness height $\lambda$. Then, an $A^*$ path detection algorithm from a
starting node SN to a target node TN was implemented\cite{Ueland2017}. The basic principles of the code are resumed by the schematic in Fig.~\ref{schematic algorithm}. For each current node $CN$, the algorithm minimises the cost function $f(N)$ to select the following node $N$:
\begin{equation}
f\left( N \right) = g\left( N \right) + h\left( N \right),
 \label{eq:cost function}
\end{equation}
where $g(N)$ is the cost based on the distance from $N$ to $TN$ and $h(N)$ is the cost based on the distance from $CN$ to $N$. The optimal path is therefore progressively evaluated from $SN$ to $TN$.

\begin{figure}[h]
\centerline{\includegraphics[width = .5 \columnwidth]{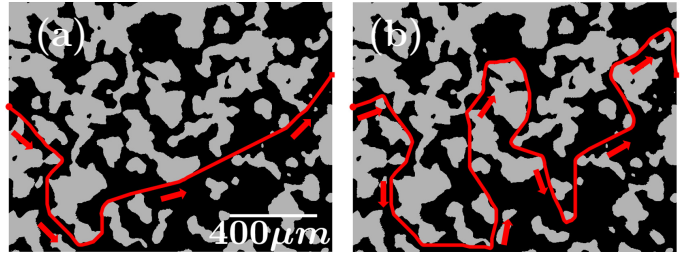}
\put(-244,78){\textbf{\textcolor{white}{\textbf{\Large(a)}}}}\put(-118,78){\textbf{\Large\textcolor{white}{(b)}}}
}
\vspace{-3mm}
\caption{Path detection on a binarised 3D confocal image of a SH-2 coating flat plate. (a) optimal path, (b) non-optimal path. The black and grey regions indicate the areas below and above the rms surface roughness $\lambda$, respectively. The red dot and square indicate the starting and target points, respectively. The red arrows indicate the path direction.}
\label{fig:3}
\end{figure}

The output of this algorithm applied to the actual SH coatings microscope images is reported in Fig.~\ref{fig:3}, where the case of the SH-2 coating is taken as an example. The value $\Theta = 1.6$ is estimated in the optimal path case in Fig.~\ref{fig:3}(a). However, by artificially forcing the code to follow non-optimal paths, the tortuosity value can quickly be increased. For example, the value $\Theta = 3.2$ is retrieved in Fig.~\ref{fig:3}(b). This shows that a precise and statistically relevant $\Theta$ estimation from the available information on the produced SH coatings is not straightforward. 
In fact, the $A^{\star}$ algorithm applied to the other SH coatings provided the same $\Theta$ order of magnitude. It is worth noting that this range corresponds to the typical empirical values reported in literature (see Bear\cite{Bear1988}, and references therein).

\begin{figure} [t!]
{\includegraphics[width = .8 \columnwidth]{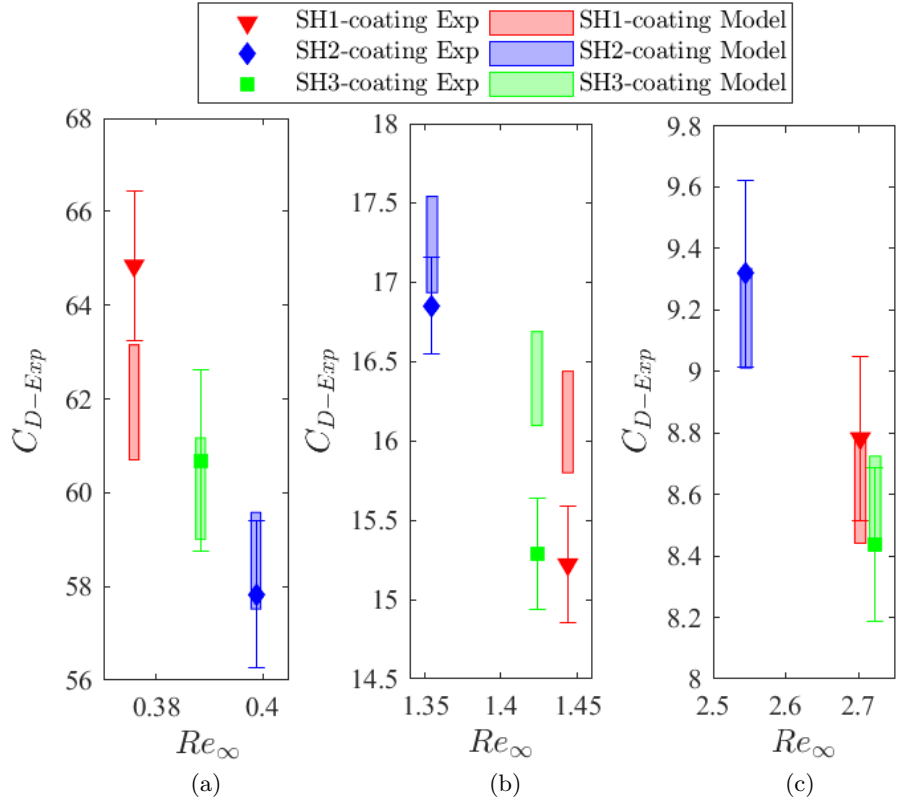}}
\put(-305,-10){(a)}
\put(-192,-10){(b)}
\put(-80,-10){(c)}
\caption{Comparison between the obtained experimental drag and that calculated using the developed model. (a) Reynolds number $Re_{\infty-1}$. (b) Reynolds number $Re_{\infty-2}$. (c) Reynolds number $Re_{\infty-3}$.
$\textcolor{red}{\blacktriangledown}$, SH-1 coating. $\textcolor{blue}{\blacklozenge}$, SH-2 coating. $\textcolor{green}{\blacksquare}$, SH-3 coating.
\begin{tikzpicture}
\fill[draw=red,fill=red!30, thin](0,0) rectangle (1,0.25);
\end{tikzpicture} predicted drag range for SH1,
\begin{tikzpicture}
\fill[draw=blue,fill=blue!30, thin](0,0) rectangle (1,0.25);
\end{tikzpicture} predicted drag range for SH2,
\begin{tikzpicture}
\fill[draw=green,fill=green!30, thin](0,0) rectangle (1,0.25);
\end{tikzpicture} predicted drag range for SH3.
}

\label{fig:barsRandomRoughness}
\end{figure}

\begin{table}[t]
\begin{ruledtabular}
\begin{tabular}{|c|cccc|cccc|cccc|}
 & $Re_\infty$ & $C_{D-NS}$ & $C_{D-Exp}$ & ${\Phi}$ & $\Theta$ & $\frac{U_{\mathrm{slip}}}{U_\infty}$ & $C_{D-SH}$ & $Error [$\%$]$&  $\Theta$&$\frac{U_{\mathrm{slip}}}{U_\infty}$& $C_{D-SH}$ & Error [$\%$]  \\ 
 \hline
\multirow{3}{4em}{SH-1} & 0.375 & 68.759 & 64.840 &\multirow{3}{2em}{0.59}& \multirow{9}{2em}{1.6}& \multirow{3}{2em}{0.0744}& 60.7& 6.3844 &\multirow{9}{2em}{3.2}& \multirow{3}{2em}{0.0168} & 63.152 & 2.6\\ 

                         & 1.443 & 19.831 & 15.220 & & & & 15.798 & 3.8 & & & 16.436 & 8 \\
                        
                         & 2.702 & 11.519 & 8.782  & & & & 8.441 & 2.436 & & & 8.782  & 0.00231 \\
                        
\multirow{3}{4em}{SH-2} & 0.398 & 64.987 & 57.832&\multirow{3}{2em}{0.54}& & \multirow{3}{2em}{0.0664}& 57.520 & 0.539 & & \multirow{3}{2em}{0.0151} & 59.579 & 3.02 \\  
                        & 1.354 & 20.989& 16.849 & & & & 16.930 & 0.484 & & & 17.536 & 4.081 \\
                        & 2.545& 12.116  & 9.317 & & & & 9.011 & 3.285 & & & 8.724 & 3.414 \\
\multirow{3}{4em}{SH-3} & 0.388 & 66.645 & 60.684 &\multirow{3}{2em}{0.55}& & \multirow{3}{2em}{0.068}& 59.002 & 2.7728&& \multirow{3}{2em}{0.154} & 61.167 & 0.795 \\  
                        & 1.423 & 20.083 & 15.286 & & & & 16.095 & 5.294 & & & 16.685 & 9.157 \\
                        & 2.722 & 11.446 & 8.436 & & & & 8.415 & 0.245 & & & 8.724 & 3.414 \\
                        
\end{tabular}
\end{ruledtabular}

\caption{Summary of computed drag, theoretically, experimentally and numerically. $Re_\infty$, Reynolds Number. $C_{D-NC}$, drag using the correlation. $C_{D-Exp}$, experimental drag. $C_{D-SH}$, computed drag using the present model.}
\label{tab:Comparison}
\end{table}

Since exact data for the tortuosity are not available, obtained values from the above described algorithm were used as limiting cases for $\Theta$. Consequently, drag was computed using two extreme values of $\Theta$ for SH spheres with randomly distributed elements ($\Theta=1.6$ and  $\Theta=3.2$). Obtained results therefore represent a maximum and minimum computed drag using the model.

Tab.~\ref{tab:Comparison} summarises the results of the computed drag for the different SH spheres. At first glance, the proposed model provides a good estimation of drag on S-H surfaces. Predicted values for the various coatings were in good agreement with experiments. The relative error for the model predictions was found to be less than approximately $9\%$. Note that the normalised slip velocity at $\Theta=1.6$ is greater than that calculated at $\Theta=3.2$. These results demonstrate that high values of the tortuosity contribute to decrease SH performances. 
The drag predicted by the model for spheres SH-1, SH-2 and SH-3 are compared to their experimental counterparts in Fig~\ref{fig:barsRandomRoughness}. Overall, the range of the predicted drag overlaps very well with the error bars from the experiments. It is assumed that the small disparity in the results stems from the inexact estimation of $\Theta$. \\

These findings emphasise the need of shorter paths for air inside the plastron such that SH-surfaces may achieve a significant drag reduction. More precisely, the air path should be approximately straight to achieve the least resistance as air flows inside the plastron. To validate this assumption, an additional SH-sphere was manufactured to investigate the effect of regularly distributed roughness elements on drag decrease. 

\subsection{Regularly distributed roughness sphere}

\begin{figure}[h]

{\includegraphics[width = .95\columnwidth]{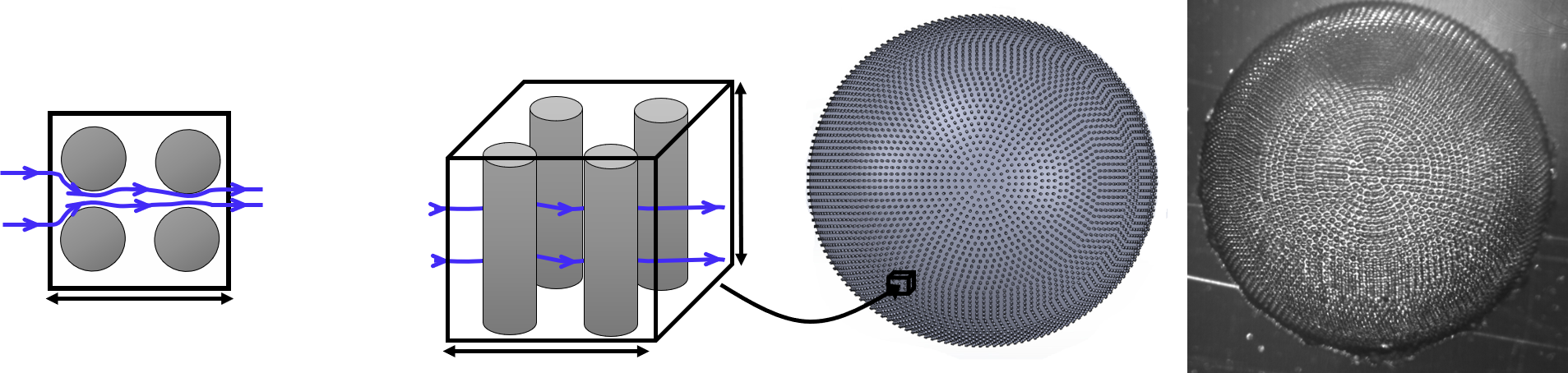}}
\put(-483,70){{\textcolor{blue}{{$L_{\Theta}$}}}}
\put(-442,10){{L}}
\put(-412,60){\textcolor{blue}{Air flow path}}
\put(-272,40){{\textcolor{blue}{{$L_{\Theta}$}}}}
\put(-318,-4){{L}}
\put(-440,-15){{Air motion inside the plastron}}
\put(-215,-15){{Digital model}}
\put(-102,-15){{3D printed SH-Sphere}}
\put(-253,57){h}

\caption{Schematic of a porous medium model over a CV (width $L$ and height $h$) inside the air layer on the surface of a SH sphere with regularly distributed pillars. The digital model of the regularly distributed roughness SH sphere was manufactured by 3D printing. Note the minimisation of the tortuosity throughout the regularly distributed roughness elements.}
\label{fig:Porous Medium-Pillars}
\end{figure}
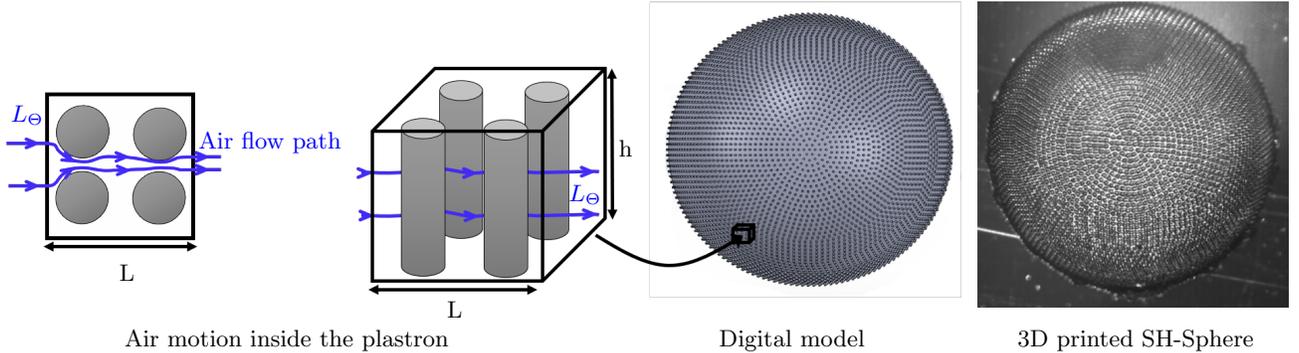

Fig.~\ref{fig:Porous Medium-Pillars} depicts a schematic of a CV with cylindrical pillars as roughness elements. Since the pillars are structurally distributed over the surface, the air flows  approximately straight ($L_{\Theta} \approx L$).  Using this concept, a SH-sphere textured with a regular distribution of pillars (SH-RDR) was 3D printed (only for the $d=10$ mm case) to assess experimentally the impact of roughness alignment on drag. Fig.~\ref{fig:Porous Medium-Pillars} also shows a 3D rendering of the model used to 3D-print an empty shell which hosts a $d=10$ mm reference stainless steel sphere. The surface roughness is characterised by aligned and equidistant cylindrical pillars with $\lambda \approx 75$ $\mu$m. The distance between pillars was also kept in the same range, once the SH coating was applied. A qualitative assessment of the presence of the air-layer around the SH sphere dipped in pure glycerine was performed to make sure that the air layer was well present. The result is presented in Fig.~\ref{fig:Porous Medium-Pillars} (3D printed SH-sphere).

Tab.~\ref{tab:Comparison SH-RDR} summarises the result for the 3D printed sphere with regular grooves and equally spaced pillars. SH-RDR shows an interesting drag reduction improvement with a mean reduction of approximately $30.7$\% with respect to the reference sphere. This outstanding performance can be explained by the fact that with pillars, the tortuosity is minimised and independent of the orientation of the sphere during its fall. This result highlights the potential role of a regular surface textures in improving drag reduction.\\

\begin{table}[t!]
\begin{ruledtabular}
\begin{tabular}{ccccccccc}
 & $Re$ & $C_{D-NS}$ & $C_{D-Exp}$ & ${\Phi}$ & $\Theta$ & $\frac{U_{\mathrm{slip}}}{U_\infty}$ & $C_{D-SH}$ & $Error [$\%$]$  \\ 

\multirow{1}{4em}{SH-RDR}& 2.439 & 12.562     & 8.7084     & 0.65& 1.0868 & 0.206 & 8.483 & 2.589 \\
                        
\end{tabular}
\end{ruledtabular}

\caption{Summary of computed drag, theoretically, experimentally and numerically for the regularly distributed roughness sphere SH-RDR. $Re$, Reynolds Number. $C_{D-NS}$, correlation drag. $C_{D-Exp}$, experimental drag. $C_{D-SH}$, computed drag using the present model.}
\label{tab:Comparison SH-RDR}
\end{table}

For the 3D printed sphere with cylindrical pillars, a relatively exact value of $\Theta$ can be calculated. The CAD model of the SH-RDR sphere was designed to display a final SH coating whose elements spacing and size were comparable with the randomly distributed SH coatings described in the previous subsection.
The regular imposed geometry also allowed an estimation of the void fraction with respect to the total volume providing a porosity ${\Phi=0.65}$. Going back to Fig.~\ref{fig:Porous Medium-Pillars}, the SH coated sphere present a relative alignment of the printed roughness elements with respect to the digital model. Consequently, the tortuosity would be slightly larger but yet very close to unity, $\Theta \gtrsim 1$. It is therefore assumed that the air motion follows a sinuous path around the pillars with an amplitude equal to half the distance between the centre of pillars. In this case an approximate tortuosity for the SH-RDR would be $\Theta=1.0868$. Tab.~\ref{tab:Comparison SH-RDR} further indicates that predicted drag for the SH-RDR sphere, which provides the best estimation of $\Theta$, is in very good agreement with the experimental value. An estimation error of $2.589\%$ was obtained. Whence, provided with exact values of $\Phi$ and $\Theta$, the present model appears to be capable of predicting accurate values of drag for SH surfaces. \\

The efficiency of the different SH coatings is further scrutinised using the following expression:
\begin{equation}
 \eta_{C_D}=\frac{{C_{D-NS}}-{C_{D-Exp}}}{{C_{D-NS}}-{C_{D-FS}}}.
\end{equation}
\begin{figure}[t!]

{\includegraphics[width = .7\columnwidth]{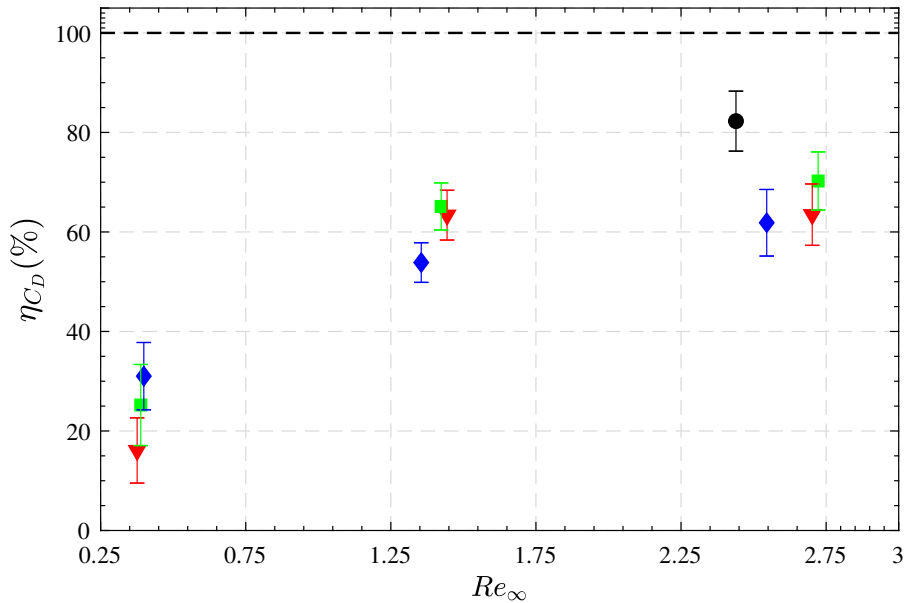}}

\caption{$\eta_{C_{D}}$ as a function of the terminal Reynolds number $Re_{\infty}$.  $\textcolor{red}{\blacktriangledown}$, SH-1 coating. $\textcolor{blue}{\blacklozenge}$, SH-2 coating. $\textcolor{green}{\blacksquare}$, SH-3 coating. $\textcolor{black}{\newmoon}$, regular roughness SH coating.
The error bars (95$\%$ confidence level) are of the same order of the size of the symbols. The dashed line
outlines the optimum performance where ${C_{D-Exp}}={C_{D-FS}}$ .}
\label{}
\end{figure}
%
%
%
%
%
%
%
%
%
%
%
%
The previous expression scales $\Delta C_{D\infty}$ between zero for the no-slip case and unity for a sphere with zero-shear stress at the wall. The only sphere that provided drag close to that obtained with a free-slip condition was the SH-RDR model whose  tortuosity is minimised and its efficiency is $82\%$. In other words, controlling the geometric properties of a SH surface may result in hydrodynamic performances close to that of a free slip condition. These findings further support the homogenisation approach as a viable mean to design and predict the performances of SH textures in the laminar regime.

\section{Conclusions}

A novel approach is introduced to elucidate the slip-reduction performances of SH surfaces at low Reynolds number where the slippery effects are found to be ascribed to the motion of entrapped air inside the surface roughness elements. This coupled fluid problem is tackled using an homogenisation approach coupling an analytical solution for the creeping flow of a liquid around a sphere and the flow of air inside complex groovy surfaces. The model is finally tested and validated against laboratory experiments with an excellent agreement.

One of the key contribution of this work is the identification of a new parameter playing an important role in the performances of SH surfaces. The lubricating effect of SH-surfaces is analysed using a porous medium approach characterised by two control parameters: the porosity and the tortuosity. While the effect of the porosity was already studied in the past\cite{ybert2007}, the effect of tortuosity is the focus of the present study.
Numerical simulations were conducted in order to compute the air flow inside the textured surface and the mean slip velocity was found to scale as power laws as a function of both the porosity and the tortuosity. These scaling laws are then used to homogenise the boundary condition for the slip velocity. The developed expressions demonstrate that the tortuosity of the domain should be minimised in order to obtain significant drag reduction. Simultaneously, the porosity of the roughness elements should remain as high as possible but should also insure the stability of the air layer (i.e. sufficiently large contact angles). The homogenised boundary condition was later implemented in the analytical model to predict drag of SH spheres manufactured using different processes.

The accuracy of the homogenised model is supported through experimental measurements of free falling super hydrophobic spheres in glycerine. Experiments were designed to highlight the particular decrease of the hydrodynamic performances of super hydrophobic surfaces due to complex surfaces with a large tortuosity. Negligible to mild drag reducing effects are reported for various randomly distributed SH coatings at low Reynolds numbers. This comparative study between the predicted values and experiments shows that the homogenised model is capable of accurately estimating drag for super hydrophobic surfaces. This in turns requires estimating values of the porosity and tortuosity and a method is provided.  

A surface texture was later proposed to test the potential role of the surface geometric properties on drag reduction. A 3D printed sphere with regular roughness elements was manufactured for this purpose. This novel sphere demonstrated promising drag reduction performances, close to free-slip conditions which confirms that tortuosity is a key parameter in improving the performances of SH coating and subsequent textures for drag-reduction applications.

To conclude, this study shows that a compromise between scalable industrial spray coating and expensive small scale regular geometry techniques should be found in the attempt to reproduce the outstanding performance of the lotus leaves on large industrial applications (see for instance Ensikat {\it et al.}\cite{Ensikat2011}). 
Finally, it must be noted that some of the conclusions proposed above could still be valid at higher Reynolds numbers, since works available in literature (see e.g. Daniello {\it et al.}\cite{Daniello2009}) have shown the pivotal role of the relative viscous sublayer thickness with respect to the super hydrophobic coating features in turbulent flows.

It is worth finally highlighting that the present model does not predict the evolution of the ratio between the detrimental surface roughness resistance and the beneficial slip lubricating at high Reynolds number which is left to future work.

\subsection*{Acknowledgements}

This work was supported by the Direction G\'en\'erale de l'Armement (DGA), Minist\`ere de la D\'efense, R\'epublique Fran\c caise and the Agence Nationale de la Recherche (ANR) through the Investissements d'Avenir Program under the Labex CAPRYSSES Project (ANR-11-LABX-0006-01).

\bibliography{biblio}

\end{document}